\documentclass[a4,11pt]{scrartcl}
\usepackage{tabularx} 
\usepackage{amsmath}  
\usepackage{graphicx} 
\usepackage[margin=1in,letterpaper]{geometry} 
\usepackage[final]{hyperref} 
\hypersetup{
	colorlinks=true,       
	linkcolor=blue,        
	citecolor=blue,        
	filecolor=magenta,     
	urlcolor=blue         
}
\usepackage{siunitx}

\usepackage{adjustbox}
\usepackage[figuresright]{rotating}

\usepackage[section]{placeins}

\usepackage{caption}
\usepackage{subcaption}

\begin{document}

\title{Transition crossing of proton beams in SIS100}
\subtitle{Longitudinal beam dynamic simulations including space charge}

\author{Thilo Egenolf\\\\
GSI Helmholtzzentrum f\"ur Schwerionenforschung GmbH, \\
Planckstrasse 1, 64291 Darmstadt, Germany}

\date{\today}
\maketitle


\begin{abstract}
The proton cycle in the SIS100 synchrotron is designed to deliver single bunches with $2\times 10^{13}$ particles at \SI{29}{\giga\electronvolt}. During the cycle beam dynamics near transition energy have to be analyzed to avoid an significant emittance blow-up. In the past two scenarios were discussed, a shift of the transition energy above the extraction energy and a transition crossing with a so called $\gamma_{\mathrm{t}}$ jump.
To estimate the possible emittance blow-up in both scenarios, longitudinal simulations are presented including the Johnsen effect based on the second order phase slip factor and the effect of longitudinal space charge. Furthermore, different parameters in the design of the $\gamma_{\mathrm{t}}$ jump are varied and a set of parameters is proposed that shows minimal longitudinal emittance growth in the simulations.
\end{abstract}


\clearpage


\section{Introduction}

In order to reach the desired SIS100 extraction energy crossing or changing the transition energy is necessary during the acceleration cycle. In the SIS100 Project Note "Overview of the Longitudinal Beam Dynamics for the SIS100 Proton Cycles" by Kornilov, Boine-Frankenheim and Ondreka (July 2013)~\cite{Kornilov2013OverviewCycles} two different scenarions are studied. Both are starting with the proton manipulations at injection energy of \SI{4}{\giga\electronvolt} (kinetic) and accelerate within \SI{650}{\milli\second} to \SI{29}{\giga\electronvolt}. However, they differ in the transition handling:
\begin{enumerate}
    \item \textbf{(Fast / Smooth) $\gamma_{\mathrm{t}}$ shift:} The lattice used for the first \SI{240}{\milli\second} has  $\gamma_{\mathrm{t}}=18.3$. After \SI{240}{\milli\second} the optics will be changed to  $\gamma_{\mathrm{t}}=45.5$, such that the beam always stays below transition during acceleration–\cite{Bar2008FAIRSIS100}. The corresponding Lorentz and phase slip factors are shown in Fig.~\ref{fig:scenario1}. The fast lattice change can, however, cause a missmatch if the acceleration ramp is not adjusted properly. The parameters of a lattice with a smooth shift in $\gamma_{\mathrm{t}}$ studied in~\cite{Sorge2012BeamEffects} is shown in Fig.~\ref{fig:scenario1b} and discussed in Sec.~\ref{sec:gamma_t_increase}.
    
    \begin{figure}[htp]
        \centering
        \hfill%
        \begin{subfigure}[t]{0.48\textwidth}
            \centering
            \includegraphics{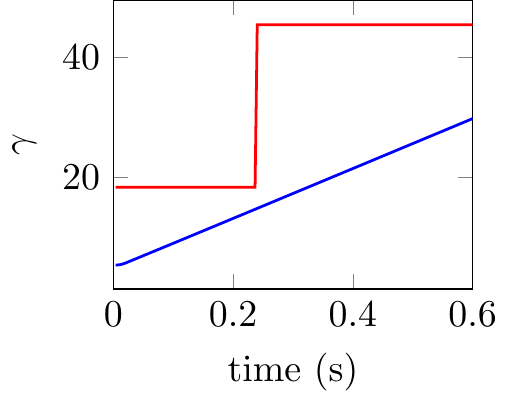}
            \caption{Acceleration ramp (blue) and $\gamma_{\mathrm{t}}$ (red)}
            \label{fig:scenario1_gamma}
        \end{subfigure}
        \hfill%
        \begin{subfigure}[t]{0.48\textwidth}
            \centering
            \includegraphics{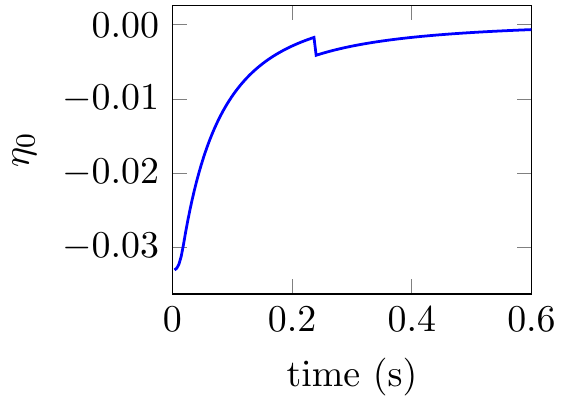}
            \caption{Phase slip factor}
            \label{fig:scenario1_eta0}
        \end{subfigure}
        \hfill%
        \caption{Parameters for scenario 1 with fast $\gamma_{\mathrm{t}}$ shift}
        \label{fig:scenario1}
    \end{figure}
    
    \begin{figure}[htp]
        \centering
        \begin{subfigure}[t]{0.32\textwidth}
            \centering
            \includegraphics{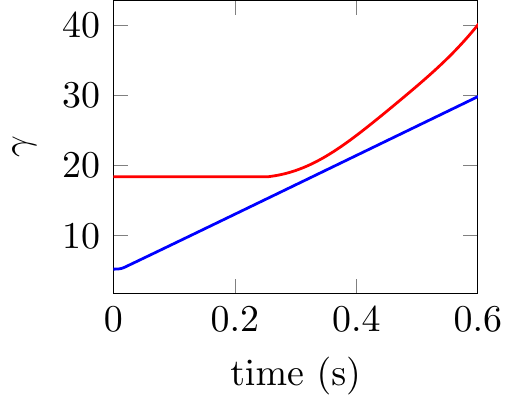}
            \caption{Acceleration ramp (blue) and $\gamma_{\mathrm{t}}$ (red)}
            \label{fig:scenario1b_gamma}
        \end{subfigure}
        \hfill
        \begin{subfigure}[t]{0.32\textwidth}
            \centering
            \includegraphics{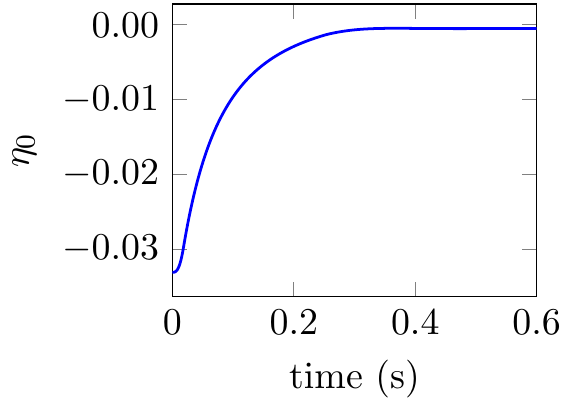}
            \caption{Phase slip factor}
            \label{fig:scenario1b_eta0}
        \end{subfigure}
        \hfill
        \begin{subfigure}[t]{0.32\textwidth}
            \centering
            \includegraphics{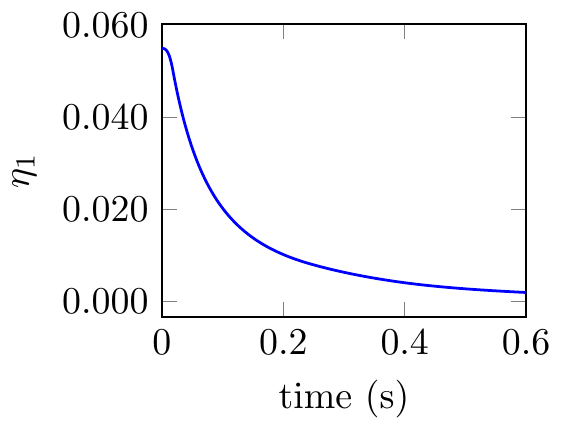}
            \caption{Second order phase slip factor}
            \label{fig:scenario1b_eta1}
        \end{subfigure}
        \caption{Parameters for scenario 1 with smooth $\gamma_{\mathrm{t}}$ shift}
        \label{fig:scenario1b}
    \end{figure}

    \item \textbf{Transition crossing / jump:} The lattice has  $\gamma_{\mathrm{t}}=8.9$ during the complete acceleration. This leads to a transition crossing along the acceleration ramp. Fig.~\ref{fig:scenario2} shows the Lorentz and phase slip factors. The simulations are discussed in Sec.~\ref{sec:transition_crossing}. In the case of a $\gamma_{\mathrm{t}}$ jump, the phase slip factors are briefly modified around transition. Why such a jump is used and the corresponding parameters are discussed in Sec.~\ref{sec:gamma_t_jump}.
    
    \begin{figure}[htp]
        \centering
        \begin{subfigure}[t]{0.32\textwidth}
            \centering
            \includegraphics{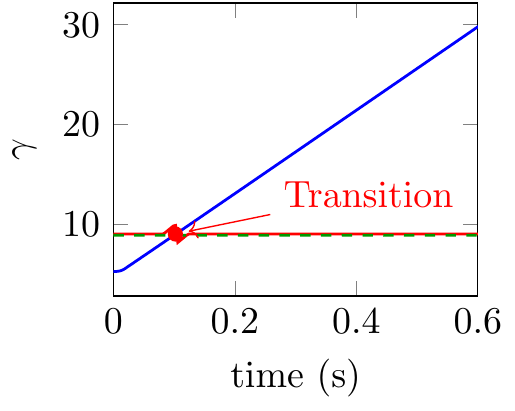}
            \caption{Acceleration ramp (blue) and $\gamma_{\mathrm{t}}$ (red with and green without jump)}
            \label{fig:scenario2_gamma}
        \end{subfigure}
        \hfill
        \begin{subfigure}[t]{0.32\textwidth}
            \centering
            \includegraphics{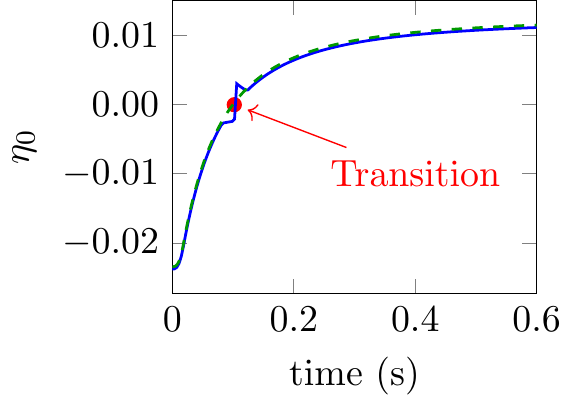}
            \caption{Phase slip factor (blue with and green without jump)}
            \label{fig:scenario2_eta0}
        \end{subfigure}
        \hfill
        \begin{subfigure}[t]{0.32\textwidth}
            \centering
            \includegraphics{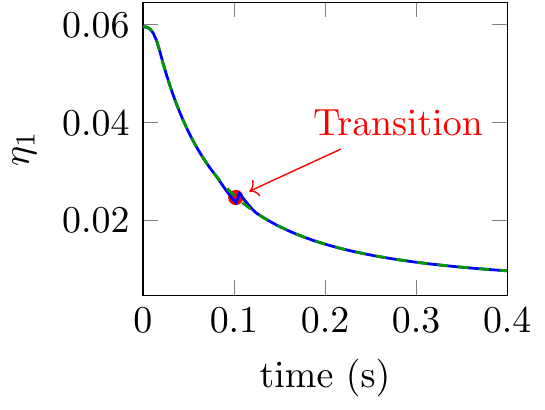}
            \caption{Second order phase slip factor (blue with and green without jump)}
            \label{fig:scenario2_eta1}
        \end{subfigure}
        \caption{Parameters for scenario 2: transition crossing / jump}
        \label{fig:scenario2}
    \end{figure}
\end{enumerate}

In the SIS100 proton cycle prior to the acceleration ramp bunch manipulations take place to merge four injected bunches to one~\cite{Yuan2021RFSIS-100}. The manipulations result in a bunch distribution which properties are listed in Tab.~\ref{tab:initial_beam_parameters}. For a maximum bucket area, the RF voltage amplitude at harmonic number $h=5$ should be as high as possible, i.e. \SI{280}{\kilo\electronvolt}, with a synchronous phase of \SI{30.46}{\deg} on the ramp. However, the matched rms momentum spread of $2.6\times10^{-3}$ for the initial emittance already exceeds the maximum momentum spread during the ramp without considering a possible increase due to the non-adiabatic behavior near transition. Therefore, the RF voltage amplitude is reduced to \SI{190}{\kilo\electronvolt} and a synchronous phase of \SI{48.34}{\deg} ensures the same energy ramp. For simulations presented in this report a simplified acceleration ramp with fixed RF voltage amplitude and synchronous phase at harmonic number $h=5$ is used (see Tab.~\ref{tab:ramp_parameters}). The constraints on the longitudinal phase space during the ramp and at final energy are given in Tab.~\ref{tab:final_beam_parameters}.

\begin{table}[thp]
\centering
\begin{minipage}[t]{0.41\linewidth}\centering
\caption{\label{tab:initial_beam_parameters} Initial parameters of proton beam after longitudinal manipulations}
\begin{tabular}{l l}
\hline
Parameter & 
Value  \\
\hline
Kinetic Energy & \SI{4}{\giga\electronvolt} \\
Intensity & $2 \times 10^{13}$ \\
Number of bunches & 1 \\
rms bunch length  &  \SI{22.6}{\metre} \\
rms momentum spread  & $1.3\times 10^{-3}$ \\
rms emittance  & \SI{0.48}{\electronvolt\second}\\\hline
\end{tabular}
\end{minipage}
\hfill%
\begin{minipage}[t]{0.57\linewidth}\centering
\caption{\label{tab:ramp_parameters} Parameters of the SIS100 acceleration ramp}
\begin{tabular}{l l}
\hline
Parameter & 
Value  \\
\hline
Magnets ramp rate & \SI{2.5}{\tesla\per\second} \\
RF voltage amplitude & \SI{280}{\kilo\volt} \\
Number of cavities  &  14 \\
Harmonic number & 5 \\
Synchronous phase below transition  & \SI{30.46}{\degree} \\
Synchronous phase above transition  & \SI{149.54}{\degree}\\
Energy change $\dot{\gamma}$ & \SI{41.09}{\per\second}\\\hline
\end{tabular}
\end{minipage}
\end{table}

\begin{table}[thp]
\begin{center}
\caption{\label{tab:final_beam_parameters} Parameters of the proton beam during and after acceleration}
\begin{tabular}{l l}
\hline
Parameter & 
Value  \\
\hline
Kinetic Energy & \SI{29}{\giga\electronvolt} \\
Maximum particle loss during ramp & $\SI{1}{\percent}$ \\
Maximum rms bunch length at final energy  &  \SI{25}{\nano\second}$\hat{=}$\SI{7.5}{\metre} \\
Maximum rms momentum spread during ramp  & $2.5\times10^{-3}$ \\\hline
\end{tabular}
\end{center}
\end{table}

\newpage
Without intensity effects the phase space dynamics near transition energy are well known in theory~\cite{Ng2006PhysicsInstabilities,Lee1999AcceleratorPhysics,Gilardoni2010Fifty1}. Since the synchrotron frequency $\omega_{\mathrm{s}}$ depends on the phase slip factor
\begin{equation}
    \eta_0 = \frac{1}{\gamma_\mathrm{t}^2}-\frac{1}{\gamma^2}\mathrm{,}
\end{equation}
it slows down if $\left|\gamma_{\mathrm{t}}-\gamma\right|\ll1$ and the adiabaticity condition $\omega_{\mathrm{s}}^{-1}\left|d\omega_{\mathrm{s}}/dt\right|\ll 1$ is not satisfied anymore. This results in a non-adiabatic synchrotron motion in a region near transition given by the nonadiabatic time~\cite{Ng2006PhysicsInstabilities}
\begin{equation}
    T_\mathrm{c}=\left(\frac{\beta_\mathrm{t}^2\gamma_\mathrm{t}^4}{2\omega_0 h}\frac{\left|\tan{\varphi_\mathrm{s}}\right|}{\dot{\gamma}^2}\right)^{1/3}
    \label{eq:nonadiabatic_time}
\end{equation}
with the synchronous phase $\varphi_{\mathrm{s}}$ and the revolution frequency $\omega_0$. A second effect on the synchrotron motion arises from the nonlinearities in the phase slip factor. Due to the high-order components of the momentum compaction factor, the phase slip factor is in general momentum spread dependent, i.e. the phase slip factor of different particles changes its sign at different times. To characterize this effect, the nonlinear time
\begin{equation}
    T_\mathrm{nl}=\frac{\gamma_\mathrm{t}\alpha_1\delta_\mathrm{max}}{2\dot{\gamma}}
    \label{eq:nonlinear_time}
\end{equation}
with $\alpha_1=\eta_1\gamma_\mathrm{t}^2$ assuming $\alpha_1=2$~\cite{Kornilov2013OverviewCycles} can be defined as the difference between the time when the phase slip factor for the synchronous particle and for the particle with momentum spread $\delta_\mathrm{max}$ changes sign. The emittance growth due to chromatic nonlinearities was first described by Johnsen~\cite{Johnsen1956EffectsTransition}. It can be estimated by~\cite{Chao1999HandbookEngineering}
\begin{equation}
    \frac{\Delta S}{S}\approx0.76\frac{T_{\mathrm{nl}}}{T_{\mathrm{c}}}
    \label{eq:johnsen_growth}
\end{equation}
with the rms bunch area $S$ if $T_{\mathrm{nl}}\ll T_{\mathrm{c}}$.

The presented simulations are done by a self-written longitudinal PIC code based on the tracking equations
\begin{eqnarray}
    \delta_j^{n+1}=\delta_j^{n}+\frac{q}{\beta_0^2 E_0} V(z_j^n) \\ 
    z_j^{n+1}=z_j^{n}-C \eta(\delta_j^{n+1}) \delta_j^{n+1}\mathrm{,}
\end{eqnarray}
where the first order phase slip factor is included by $\eta(\delta) = \eta_0 + \eta_1 \delta$, $\beta_0$ and $E_0=\gamma_0 m c^2$ are speed and energy of the synchronous particle, respectively, and the voltage potential $V(z_j^n)$ is given as a sum of the rf potential and the space charge potential.

\section{Acceleration without transition crossing: \texorpdfstring{$\gamma_{\mathrm{t}}$}{gamma transition} shift\label{sec:gamma_t_increase}}
\begin{figure}
        \centering
        \begin{subfigure}[t]{0.9\textwidth}
            \centering
            \includegraphics{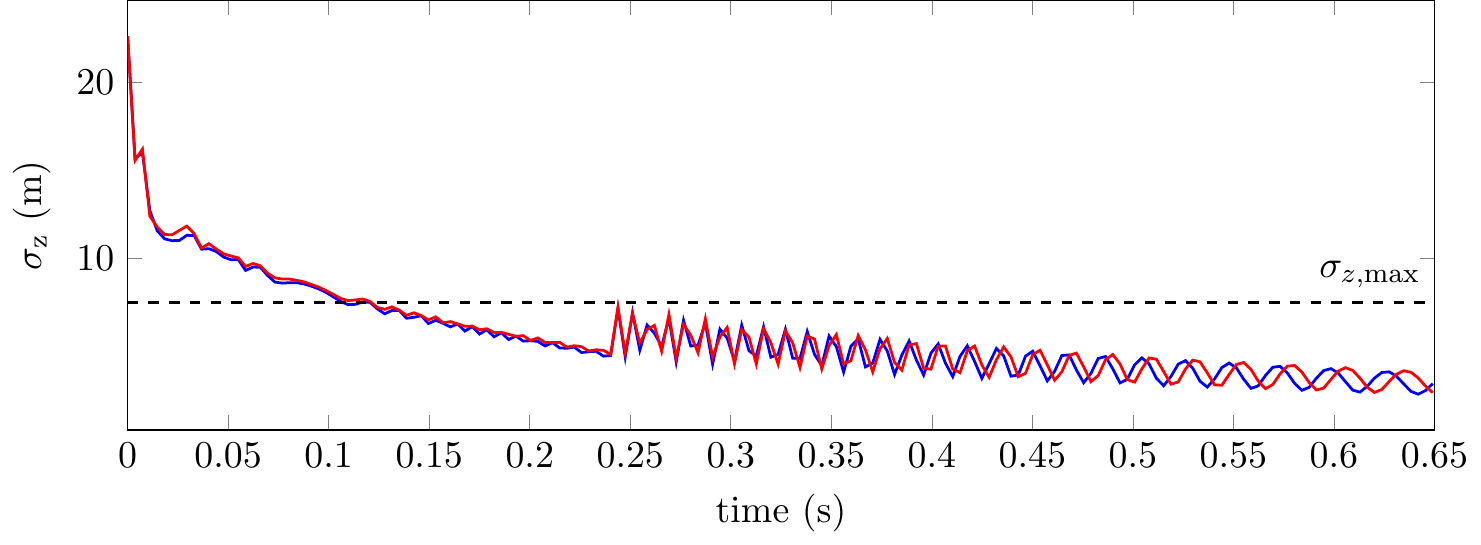}
            \caption{RMS bunch length}
            \label{fig:scenario1_bunch_length}
        \end{subfigure}
        \hfill
        \begin{subfigure}[t]{0.9\textwidth}
            \centering
            \includegraphics{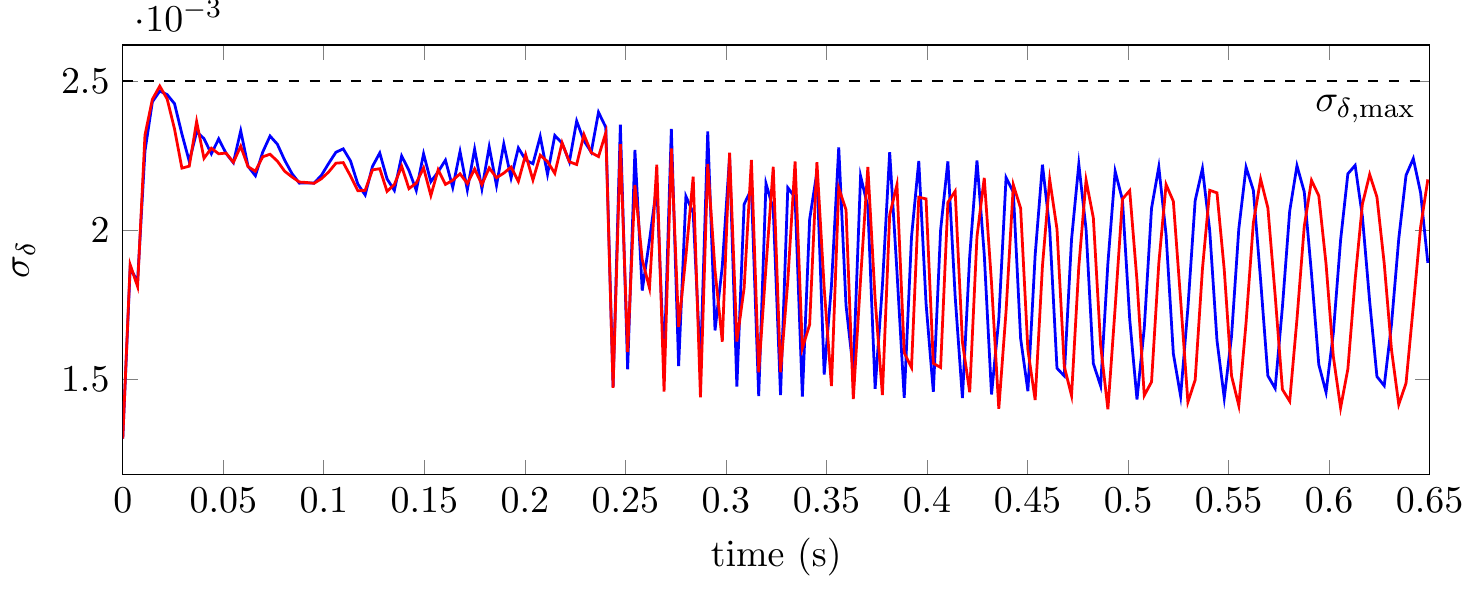}
            \caption{RMS momentum spread}
            \label{fig:scenario1_momentum spread}
        \end{subfigure}
        \hfill
        \begin{subfigure}[t]{0.9\textwidth}
            \centering
            \includegraphics{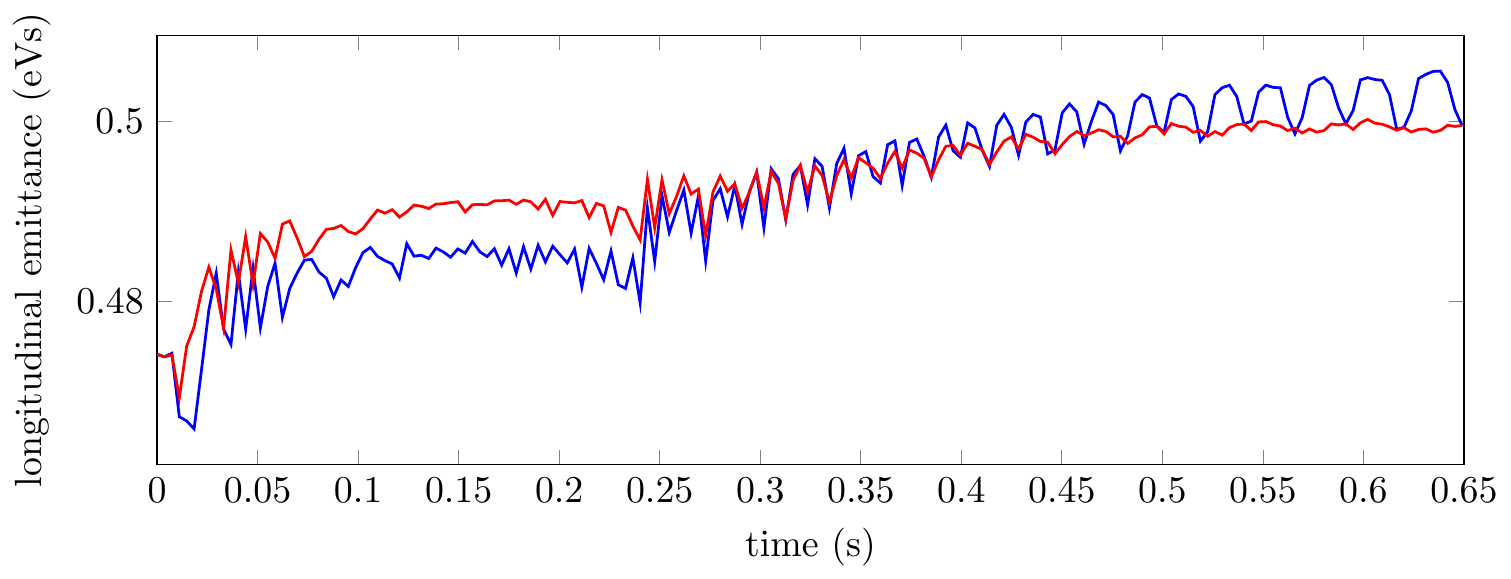}
            \caption{Longitudinal rms emittance}
            \label{fig:scenario1_emittance}
        \end{subfigure}
        \caption{Statistics for scenario 1 with fast $\gamma_{\mathrm{t}}$ increase leading to a missmatch in the bunch distribution. Results of simulations with parameters plotted in Fig.~\ref{fig:scenario1} are shown in blue and red without and with space charge, respectively.}
        \label{fig:scenario1_statistics}
    \end{figure}

To avoid transition crossing during the acceleration ramp, the first scenario includes additional magnets in the SIS100 lattice used to increase the transition energy within the cycle. In the simulations the lattice change is implemented as a discontinuous phase slip factor (see Fig.~\ref{fig:scenario1_eta0}). The initial distribution is tracked through the acceleration ramp lasting for \SI{650}{\milli\second} or 179000 turns and the resulting statistics are plotted in Fig.~\ref{fig:scenario1_statistics}. This shows that the longitudinal emittance is almost conserved during acceleration and the distribution meets the requirements for the accelerated SIS100 bunch (see Tab.~\ref{tab:final_beam_parameters}). However, the sudden change of transition energy leads to a mismatch and therefore oscillations in the longitudinal phase space larger than the effect of space charge on the bunch distribution. A more suitable acceleration ramp would probably allow the distribution in phase space to be matched to the requirements after the lattice change so that the oscillations are avoided. 

\begin{figure}
        \centering
        \begin{subfigure}[t]{0.9\textwidth}
            \centering
            \includegraphics{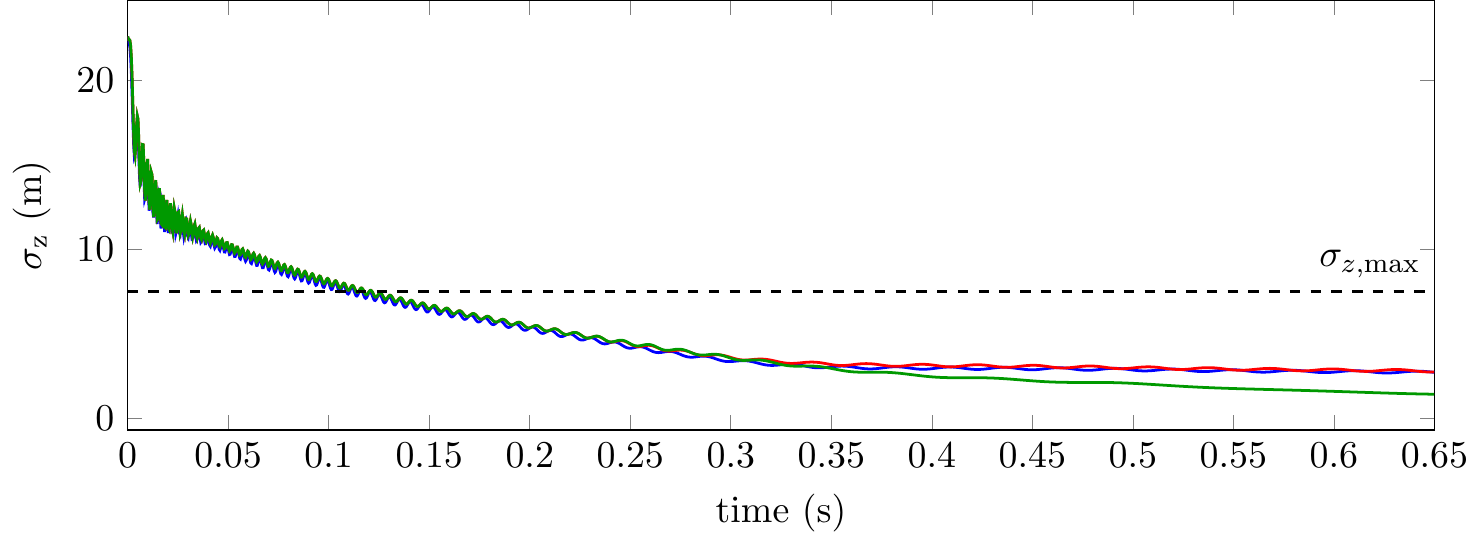}
            \caption{RMS bunch length}
            \label{fig:scenario1b_bunch_length}
        \end{subfigure}
        \hfill
        \begin{subfigure}[t]{0.9\textwidth}
            \centering
            \includegraphics{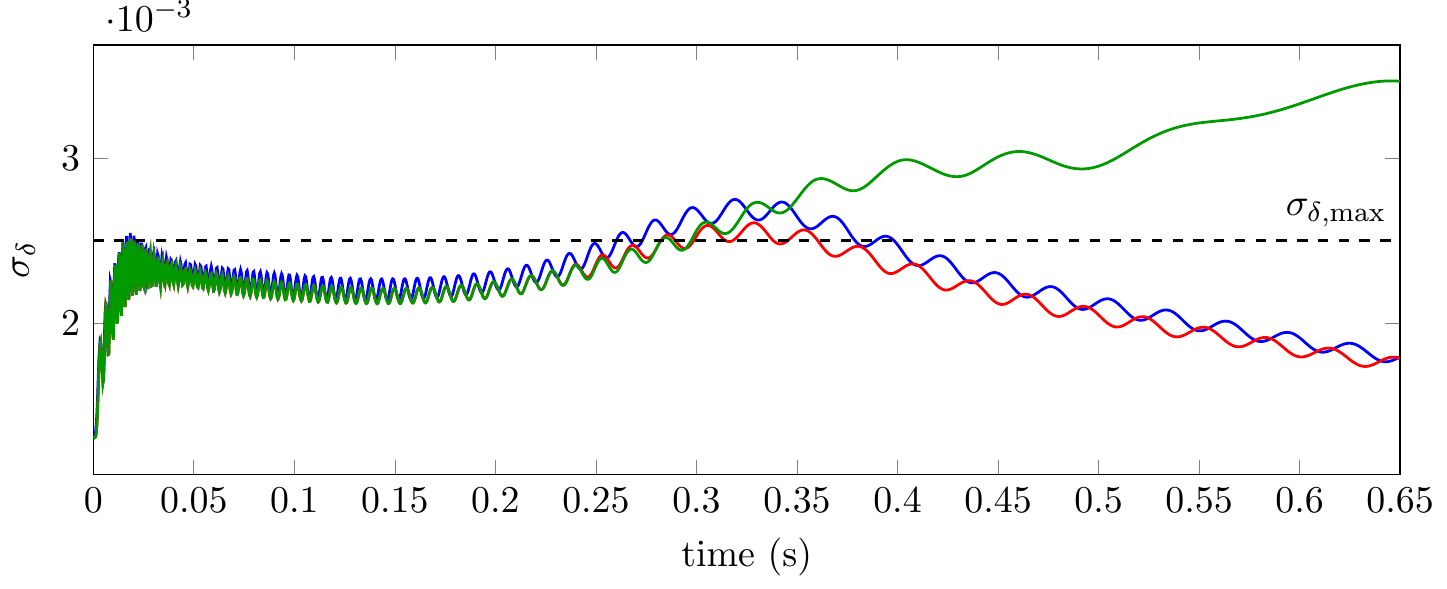}
            \caption{RMS momentum spread}
            \label{fig:scenario1b_momentum spread}
        \end{subfigure}
        \hfill
        \begin{subfigure}[t]{0.9\textwidth}
            \centering
            \includegraphics{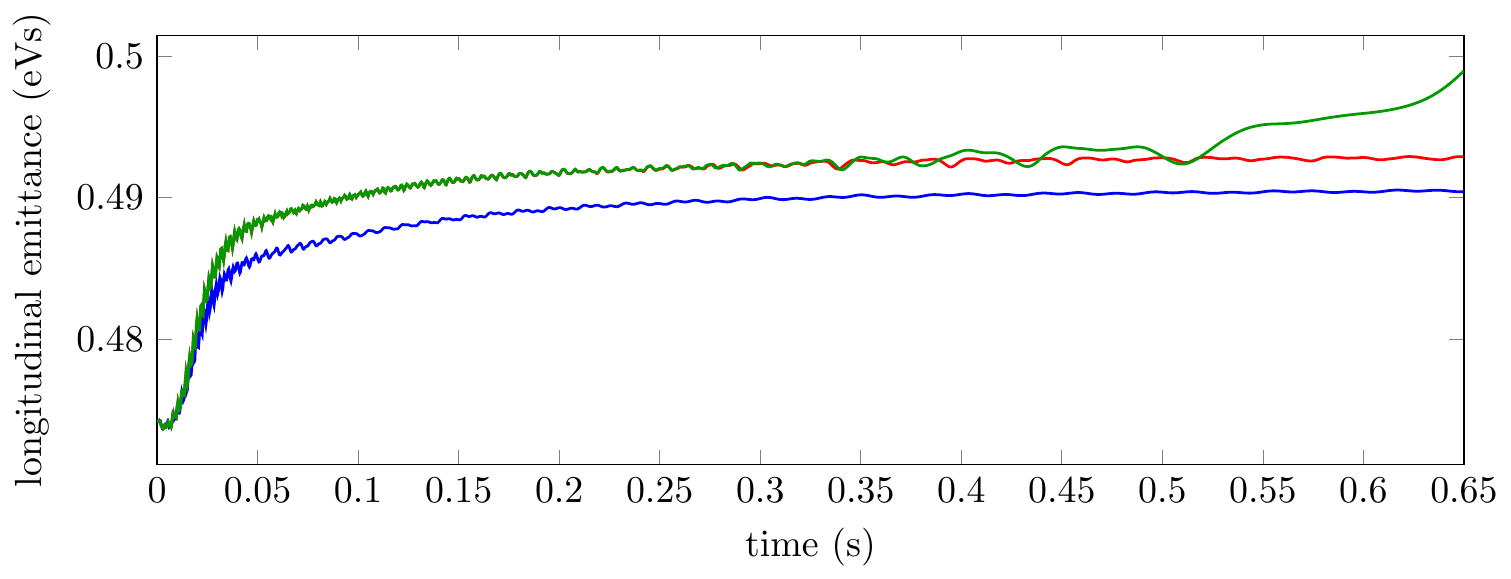}
            \caption{Longitudinal rms emittance}
            \label{fig:scenario1b_emittance}
        \end{subfigure}
        \caption{Statistics for scenario 1 with smooth $\gamma_{\mathrm{t}}$ increase. Results of simulations with parameters plotted in Fig.~\ref{fig:scenario1b} are shown in blue and red without and with space charge, respectively. The green curves shows the statistics for minimal bunch length by keeping transition energy as low as possible (cf. Fig.~\ref{fig:scenario1c}).}
        \label{fig:scenario1b_statistics}
    \end{figure}

    Another option for better matching is a smooth change of the transition energy along the acceleration ramp (cf. Fig.~\ref{fig:scenario1b}). This was proposed by S. Sorge in 2012~\cite{Sorge2012BeamEffects}. It causes no additional oscillations during acceleration and leads to an almost perfect longitudinal emittance conversation (see Fig.~\ref{fig:scenario1b_statistics}). Only small quadrupolar oscillations due to a small missmatch at the beginning of acceleration remains during the ramp. A better chosen RF voltage ramp at the beginning or a longitudinal feedback system~\cite{Lens2013StabilitySynchrotrons} could cure that. 
    
\begin{figure}[htp]
        \centering
        \begin{subfigure}[t]{0.32\textwidth}
            \centering
            \includegraphics{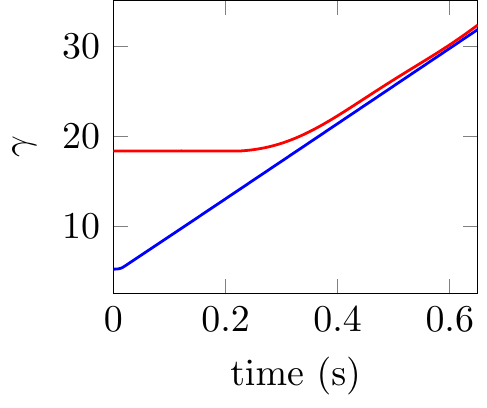}
            \caption{Acceleration ramp (blue) and $\gamma_{\mathrm{t}}$ (red)}
            \label{fig:scenario1c_gamma}
        \end{subfigure}
        \hfill
        \begin{subfigure}[t]{0.32\textwidth}
            \centering
            \includegraphics{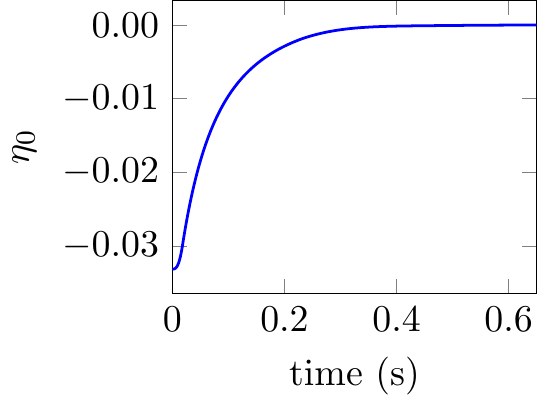}
            \caption{Phase slip factor}
            \label{fig:scenario1c_eta0}
        \end{subfigure}
        \hfill
        \begin{subfigure}[t]{0.32\textwidth}
            \centering
            \includegraphics{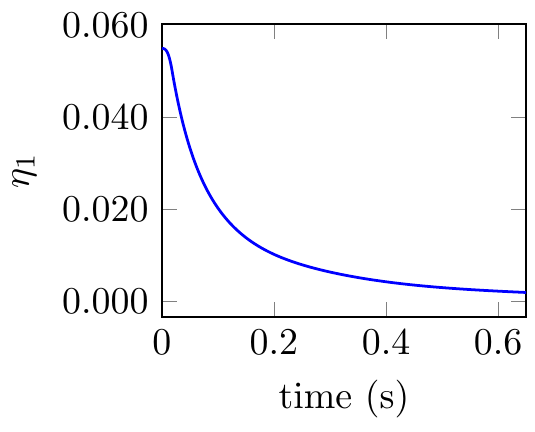}
            \caption{Second order phase slip factor}
            \label{fig:scenario1c_eta1}
        \end{subfigure}
        \caption{Parameters for scenario 1 with smooth $\gamma_{\mathrm{t}}$ increase but keeping transition energy as low as possible for minimal bunch length}
        \label{fig:scenario1c}
    \end{figure}

\begin{figure}[htp]
        \centering
        \includegraphics{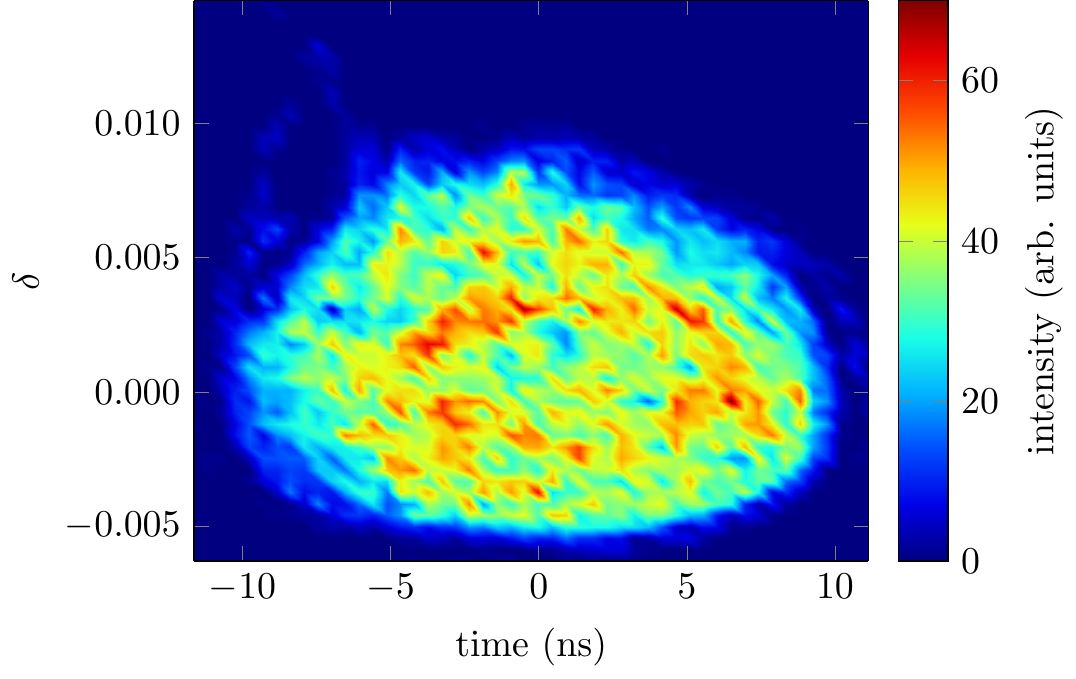}
        \caption{Phase space at the end of the acceleration ramp for scenario 1 with smooth $\gamma_{\mathrm{t}}$ increase but keeping transition energy as low as possible for minimal bunch length (cf. parameters in Fig.~\ref{fig:scenario1c} and statistics in Fig.~\ref{fig:scenario1b_statistics}). The asymmetry in momentum spread indicates that the dymanics are located in the alpha bucket regime~\cite{Sorge2018SimulationOperation}.}
        \label{fig:scenario1c_phasespace}
    \end{figure}

Although the proposed parameters provide a suitable longitudinal phase space at flat-top, reducing the proposed increase in transition energy would lead to a smaller bunch length but larger momentum spread at extraction energy. The minimum bunch length is limited by the momentum acceptance during the ramp. At the same time, enough distance between the energy of the synchronous particle and the transition energy has to be kept to avoid an emittance blow-up due to nonlinear dynamics. A trade-off between minimum bunch length, minimum emittance growth and keeping the momentum acceptance is found in simulations by varying the time dependent function of the transition energy (cf. Fig.~\ref{fig:scenario1c}). This results in a bunch with $\sigma_z\approx\SI{1.4}{\metre}=\SI{5}{\nano\second}$ and $\sigma_\delta\approx4.5\times10^{-3}$ at flat-top (see green curves in Fig.~\ref{fig:scenario1b_statistics} and phase space in Fig~\ref{fig:scenario1c_phasespace}. The asymmetric bunch shape in momentum spread indicates that the beam dynamics are located in the alpha bucket regime since the zero order phase slip factor is very close to zero (cf.~\cite{Sorge2018SimulationOperation}).

\section{Transition crossing\label{sec:transition_crossing}}

\begin{figure}
        \centering
        \begin{subfigure}[t]{0.9\textwidth}
            \centering
            \includegraphics{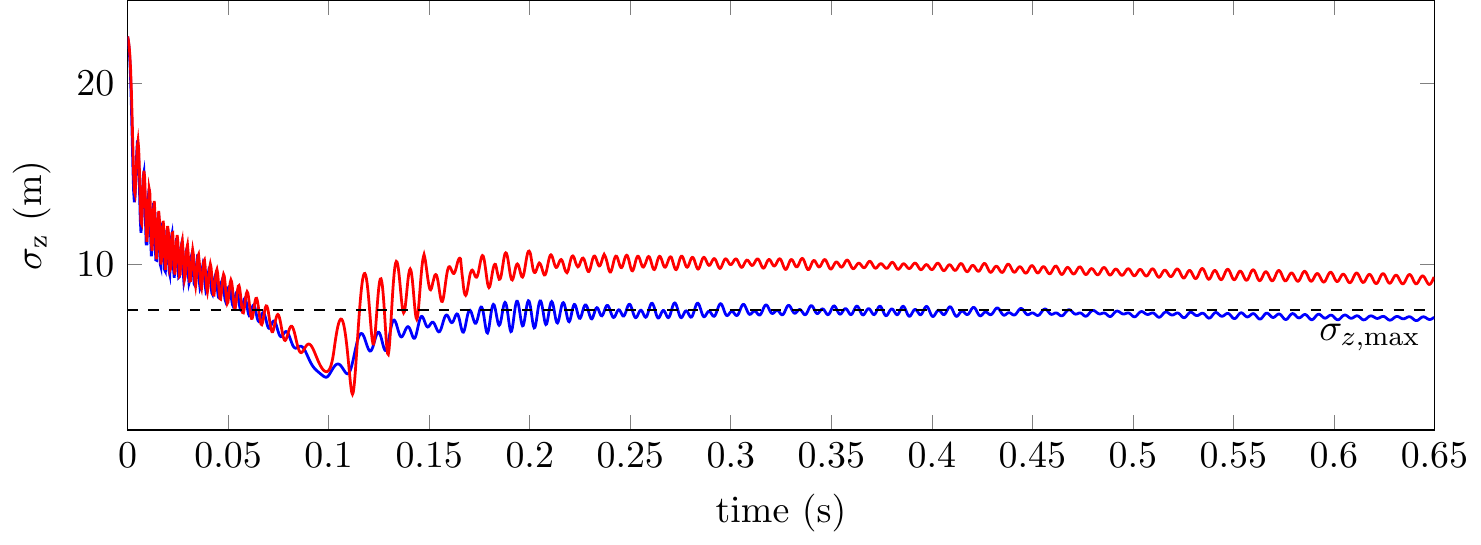}
            \caption{RMS bunch length}
            \label{fig:scenario2_bunch_length}
        \end{subfigure}
        \hfill
        \begin{subfigure}[t]{0.9\textwidth}
            \centering
            \includegraphics{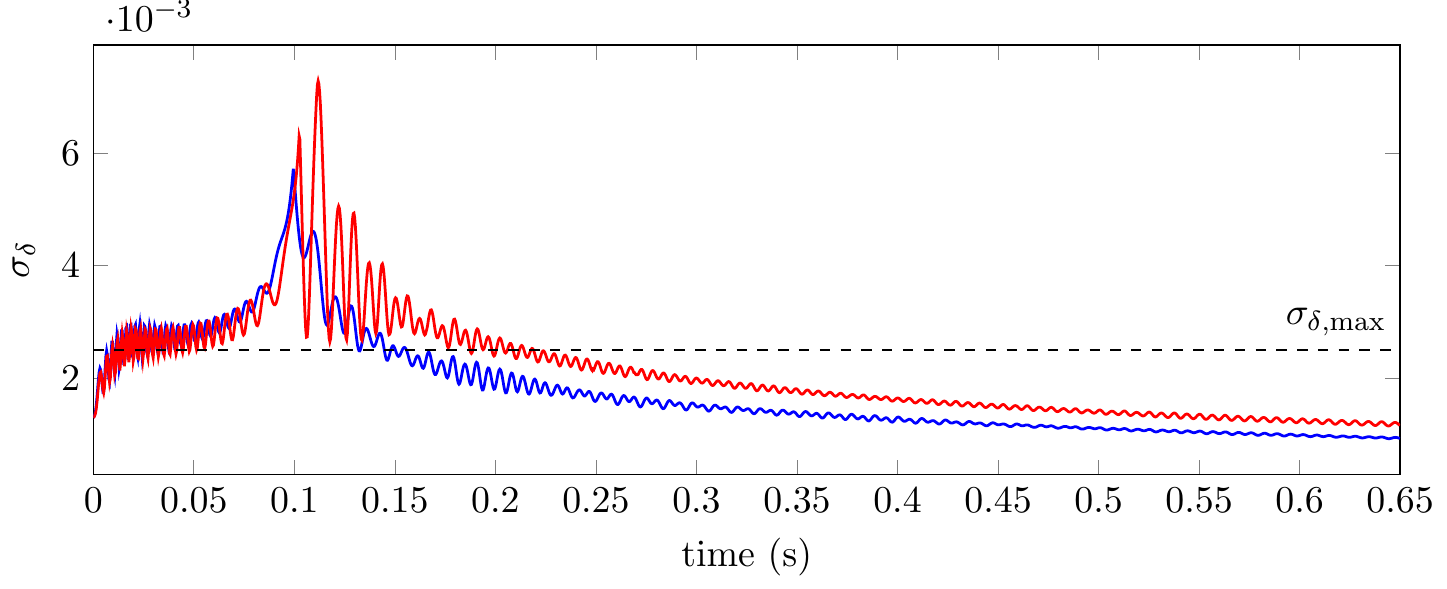}
            \caption{RMS momentum spread}
            \label{fig:scenario2_momentum spread}
        \end{subfigure}
        \hfill
        \begin{subfigure}[t]{0.9\textwidth}
            \centering
            \includegraphics{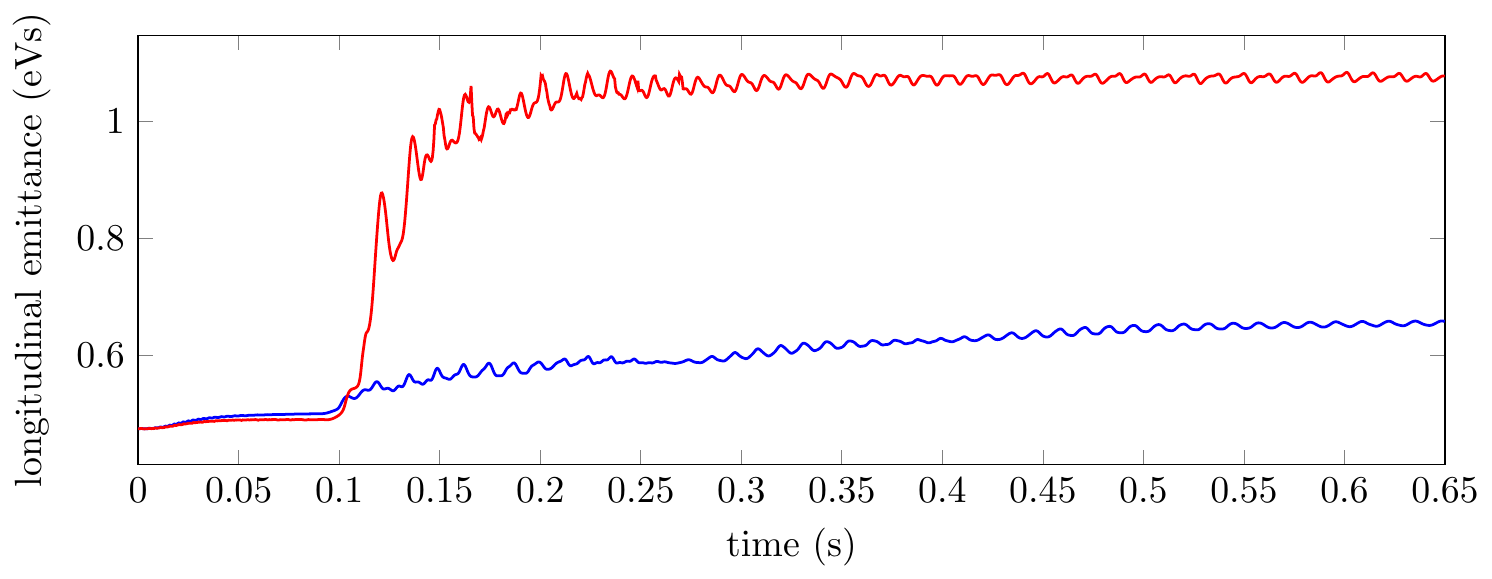}
            \caption{Longitudinal rms emittance}
            \label{fig:scenario2_emittance}
        \end{subfigure}
        \caption{Statistics for scenario 2: transition crossing without any cure for possible emittance blow-up. Results of simulations with parameters plotted in Fig.~\ref{fig:scenario2} are shown in blue and red without and with space charge, respectively.}
        \label{fig:scenario2_statistics}
    \end{figure}

The second scenario would omit additional magnets, but contains a transition crossing after \SI{102}{\milli\second} (see Fig.~\ref{fig:scenario2}). Fig.~\ref{fig:scenario2_statistics} shows the bunch length, momentum spread and longitudinal emittance along the ramp. As predicted in~\cite{Kornilov2013OverviewCycles} the transition crossing is clearly seen by the minimum of the bunch length and maximum of the momentum spread, that even exceeds the maximum possible value without particle loss (cf. Tab.~\ref{tab:final_beam_parameters}).  The nonadiabatic time around transition crossing (see Eq.~\ref{eq:nonadiabatic_time}) is $T_{\mathrm{c}}\approx\SI{5}{\milli\second}$. Within this time there is a sudden emittance growth in both simulations without and with space charge followed by a slight increase afterwards. At the extraction energy of \SI{29}{\giga\electronvolt} the bunch distribution without space charge has a rms bunch length of $\sigma_{\mathrm{z}}\approx\SI{7.1}{\metre}=\SI{23.7}{\nano\second}$, a momentum spread of $\sigma_\delta\approx9.3\times10^{-4}$ and a longitudinal emittance of $\varepsilon_{\mathrm{z}}\approx\SI{0.66}{\electronvolt\second}$, i.e. an increase of almost \SI{40}{\percent} during the acceleration ramp -- mainly caused by the transition crossing. This can be compared to the theory: With the maximum momentum spread of about $\delta_{\mathrm{max}}=5\times10^{-3}$ the nonlinear time (Eq.~\ref{eq:nonlinear_time}) in this scenario is $T_{\mathrm{nl}}\approx\SI{1.1}{\milli\second}$. The emittance growth by the Johnsen effect (Eq.~\ref{eq:johnsen_growth}) is thus $\Delta S/S\approx\SI{16.7}{\percent}$ which corresponds very well to the sudden emittance growth at transition.  However, with space charge the emittance at extraction additionally increases by \SI{162}{\percent} to $\varepsilon_{\mathrm{z}}\approx\SI{1.07}{\electronvolt\second}$, the rms bunch length to $\sigma_{\mathrm{z}}\approx\SI{9.31}{\metre}=\SI{31}{\nano\second}$ and the momentum spread to $\sigma_\delta\approx11.6\times10^{-4}$, such that the final bunch length is larger than the maximum tolerable value (cf. Tab.~\ref{tab:final_beam_parameters}). Furthermore, the momentum spread exceeds the maximum tolerable value slightly at the beginning of the ramp and significantly around transition, although beam loading and other impedance sources as possible candidates for additional longitudinal emittance growth are not yet included in the presented simulations. To keep the momentum spread below the limit, the acceleration ramp is modified in the next section and a jump of the transition energy is introduced in the section after next. 

    \section{Transition crossing with modified acceleration ramp\label{sec:modified_ramp}}
\begin{figure}[htp]
        \centering
        \includegraphics{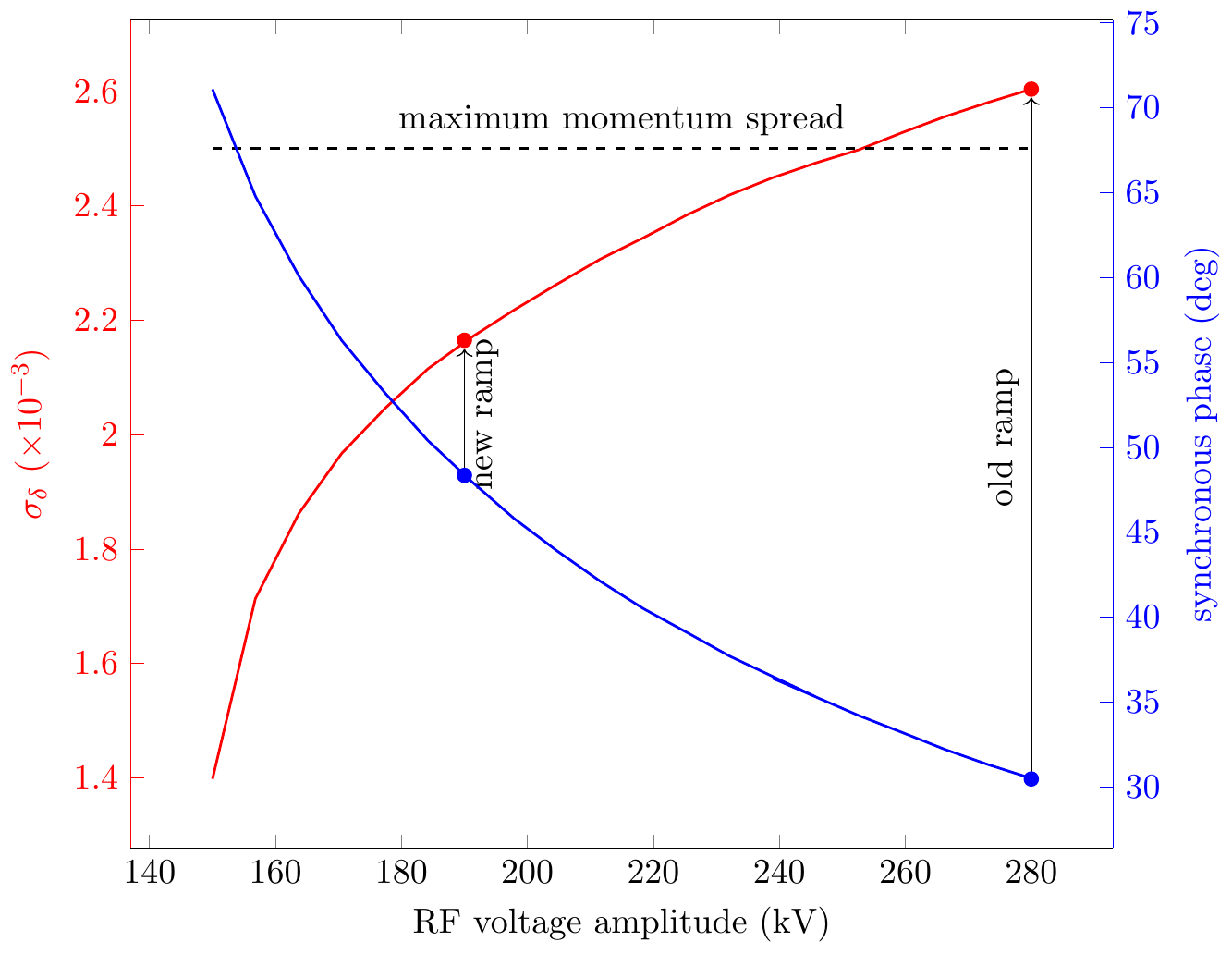}
        \caption{Synchronous phase (blue) and matched momentum spread (red) for a given RF voltage amplitude to match the initial longitudinal emittance and energy change per turn (cf. Tabs.~\ref{tab:initial_beam_parameters} and~\ref{tab:ramp_parameters})}
        \label{fig:sync_phase_momentum_spread}
    \end{figure}

    For the previous sections the acceleration ramp is chosen such that the desired energy change per turn is reached by the maximum available RF voltage amplitude resulting in the largest bucket area (cf. Tab.~\ref{tab:ramp_parameters}). However, the matched momentum spread for the given initial emittance in the transition crossing scenario (see Fig.~\ref{fig:scenario2}) is $\sigma_\delta=2.6\times10^{-3}$, i.e., above the limit given in Tab.~\ref{tab:ramp_parameters}. If the RF voltage amplitude is decreased and the synchronous phase is increased simultaneously to gain the desired energy per turn (see Fig.~\ref{fig:sync_phase_momentum_spread}), the matched momentum spread also decreases. As new initial ramp parameters a RF voltage amplitude of $\SI{190}{\kilo\volt}$ and a synchronous phase of \SI{48.35}{\degree} are used resulting in a matched momentum spread of $\sigma_\delta=2.17\times10^{-3}$ well below the maximum value but at the same time in a bucket area which is large enough to accelerate the bunch without particle loss.

    \begin{figure}[htp]
        \centering
        \includegraphics{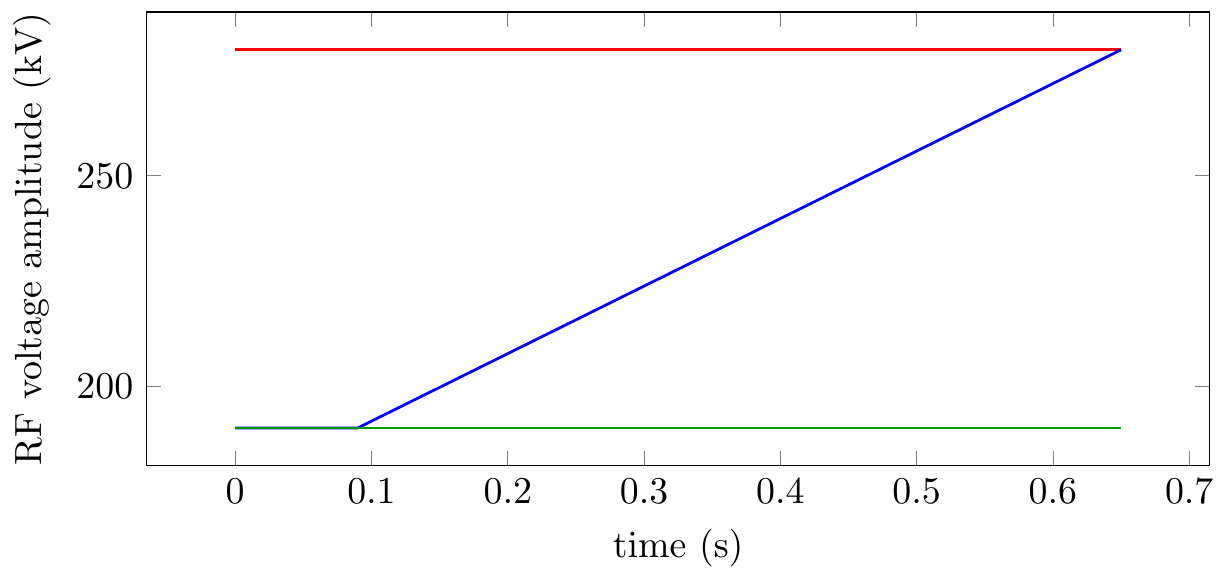}
        \includegraphics{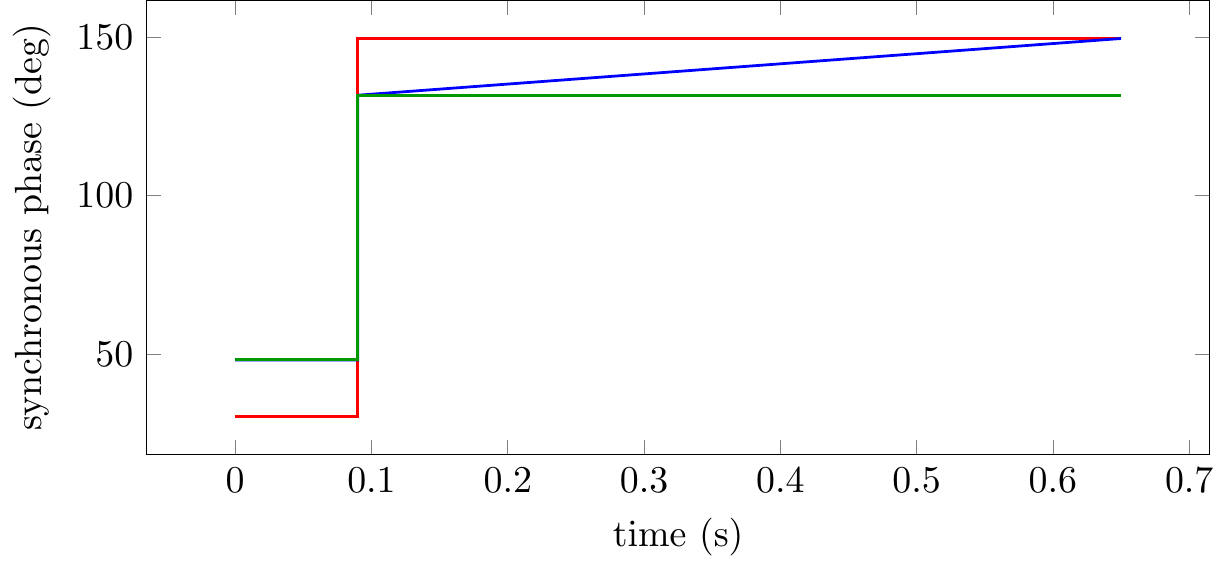}
        \caption{RF voltage amplitude (top) and synchronous phase (bottom) of the different ramps: the initial ramp with constant amplitude and phase  (red), the modified ramp with decreased but constant amplitude and increased but constant phase (blue) and the mixed one with constant amplitude and phase below transition and linearly changing parameters above transition (green).}
        \label{fig:modified ramp}
    \end{figure}

\begin{figure}
        \centering
        \begin{subfigure}[t]{0.9\textwidth}
            \centering
            \includegraphics{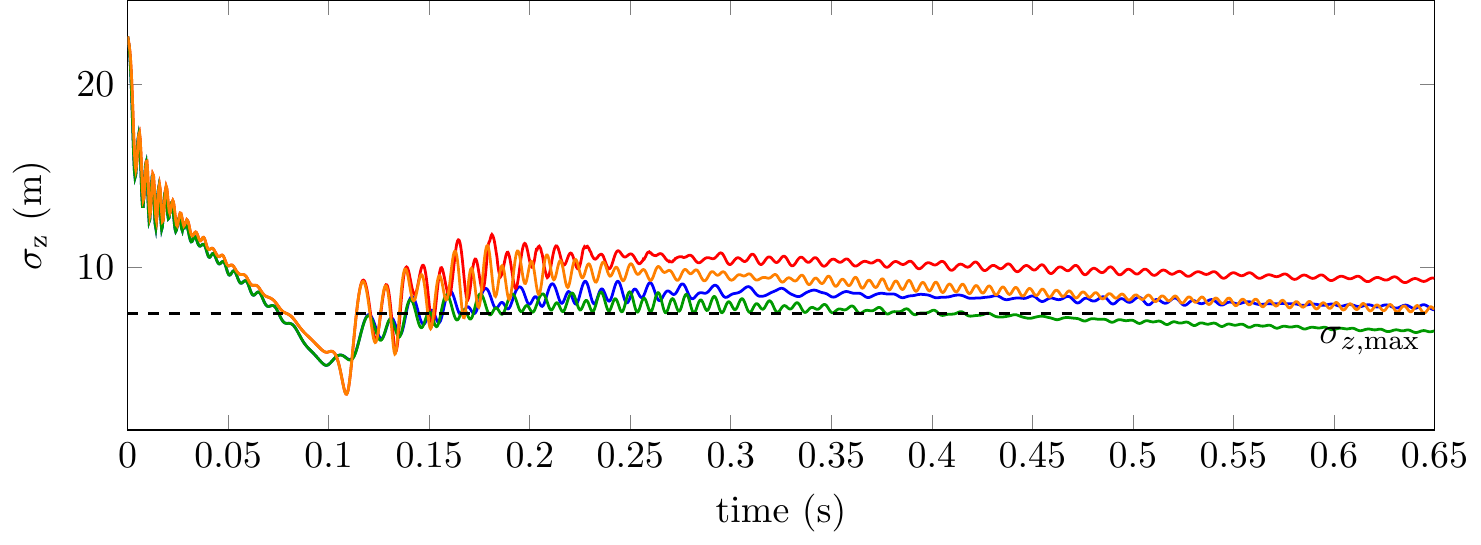}
            \caption{RMS bunch length}
            \label{fig:scenario2_modified_bunch_length}
        \end{subfigure}
        \hfill
        \begin{subfigure}[t]{0.9\textwidth}
            \centering
            \includegraphics{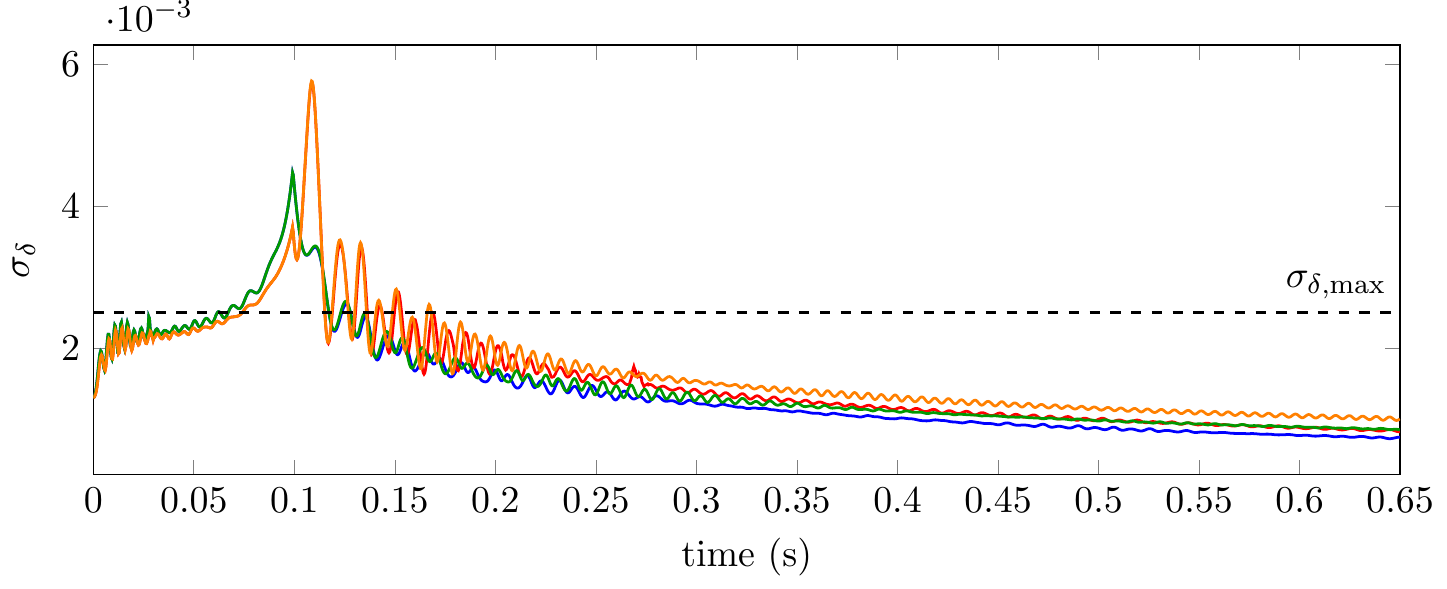}
            \caption{RMS momentum spread}
            \label{fig:scenario2_modified_momentum spread}
        \end{subfigure}
        \hfill
        \begin{subfigure}[t]{0.9\textwidth}
            \centering
            \includegraphics{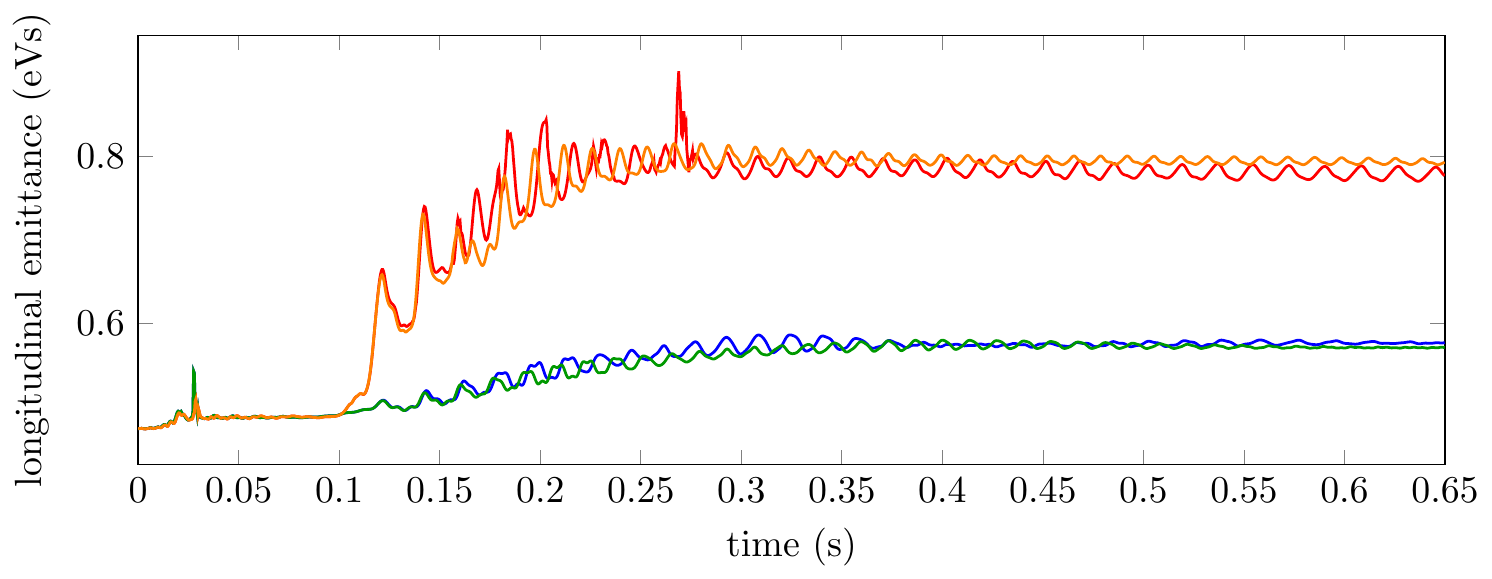}
            \caption{Longitudinal rms emittance}
            \label{fig:scenario2_modified_emittance}
        \end{subfigure}
        \caption{Statistics for scenario 2 with modified acceleration ramp: transition crossing without any cure for possible emittance blow-up. Results of simulations with parameters plotted in Fig.~\ref{fig:scenario2}, \SI{190}{\kilo\volt} RF amplitude and a synchronous phase of \SI{48.35}{\degree} are shown in blue and red without and with space charge, respectively. The green (without space charge) and orange (with space charge) lines show the simulation results if the RF voltage amplitude increases linearly to \SI{280}{\kilo\volt} after transition and the synchronous phase decreases accordingly.}
        \label{fig:scenario2_modified_statistics}
    \end{figure}

    The simulation results with the modified RF voltage and synchronous phase are shown in Fig.~\ref{fig:scenario2_modified_statistics}. As expected, the momentum spread stays below the maximum value before transition crossing. However, the momentum spread exceeds the limit near to transition, the final bunch length is too large, especially with space charge, and there is still a significant emittance growth at transition in the simulation with space charge. With a linearly increasing RF voltage amplitude up to \SI{280}{\kilo\volt} and a accordingly decreasing synchronous phase, the final bunch length is almost exactly at the maximum value but of course this can not prevent the peak in momentum spread and the emittance growth at transition. Therefore, a $\gamma_{\mathrm{t}}$ jump is necessary.

    
\section{Transition crossing with \texorpdfstring{$\gamma_\mathrm{t}$}{gamma transition} jump\label{sec:gamma_t_jump}}
\begin{figure}[htp]
        \centering
        \includegraphics{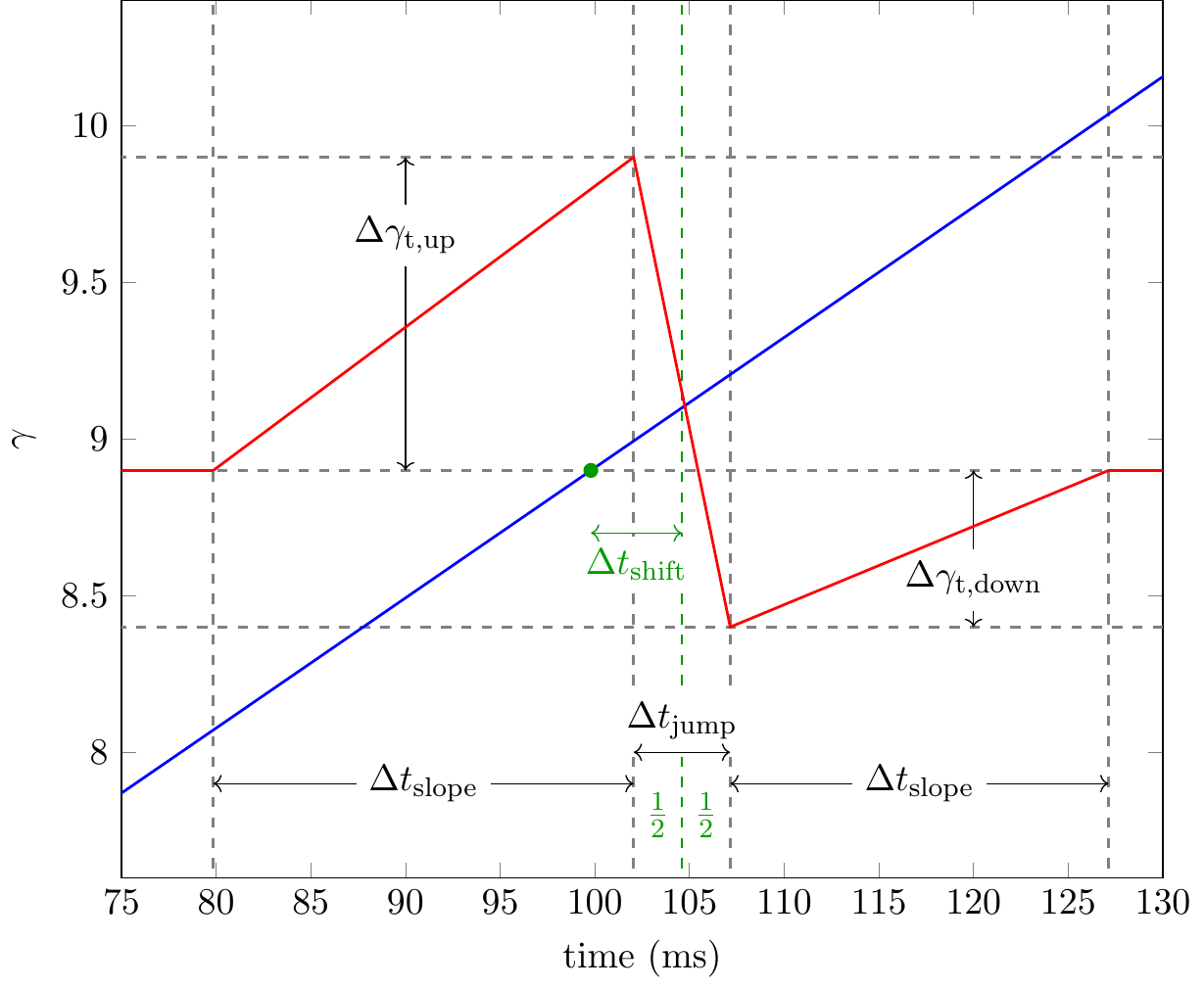}
        \caption{Jump of the transition energy around transition crossing}
        \label{fig:gamma_transition_jump}
    \end{figure}

Crossing transition at planned SIS100 intensity leads to a not tolerable emittance growth and bunch lengthening due to nonlinear effects and space charge. To prevent this, a jump of the transition energy in the range of the nonadiabatic time can be introduced. For this, the transition energy is increased before the particle energy reaches transition, rapidly decreased below the initial transition energy and then again increased to stay at the initial transition energy for the remaining acceleration above transition. Fig.~\ref{fig:gamma_transition_jump} shows an example of such a $\gamma_{\mathrm{t}}$ jump. The jump can be described by a shift $\Delta t_{\mathrm{shift}}$ with respect to the original transition crossing, the length $\Delta t_{\mathrm{jump}}$ between maximum and minimum of $\gamma_{\mathrm{t}}$, the values $\Delta\gamma_\mathrm{t,up/down}$ of these two extrema and the increasing and decreasing slopes $\Delta t_{\mathrm{slope}}$ in front of the maximum and after the minimum. 

    In the following, the emittance growth caused by the $\gamma_{\mathrm{t}}$ jump, the maximum momentum spread (during the jump) and the final bunch length after the ramp are studied by varying these jump parameters.
    
    \begin{figure}
        \centering
        \begin{subfigure}[t]{0.32\textwidth}
            \centering
            \includegraphics{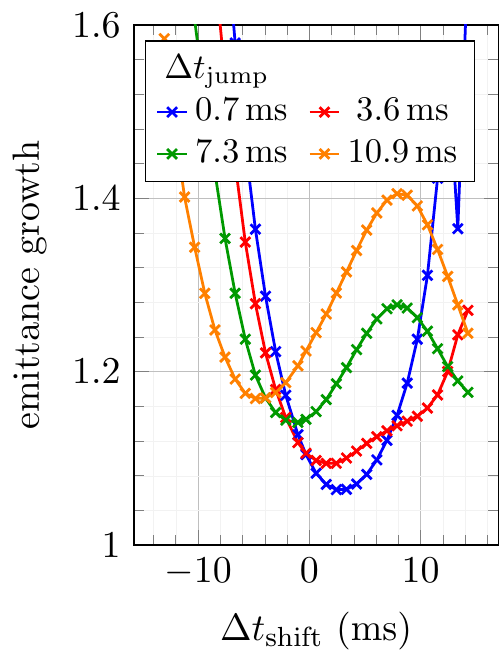}
            \caption{$\Delta\gamma_{\mathrm{t,up}}=\Delta\gamma_{\mathrm{t,down}}=0.5$}
        \end{subfigure}
        \hfill
        \begin{subfigure}[t]{0.32\textwidth}
            \centering
            \includegraphics{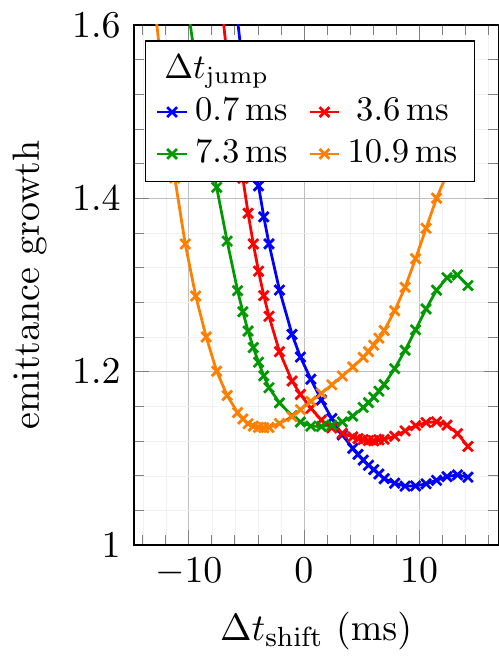}
            \caption{$\Delta\gamma_{\mathrm{t,up}}=1.0$, $\Delta\gamma_{\mathrm{t,down}}=0.5$}
        \end{subfigure}
        \hfill
        \begin{subfigure}[t]{0.32\textwidth}
            \centering
            \includegraphics{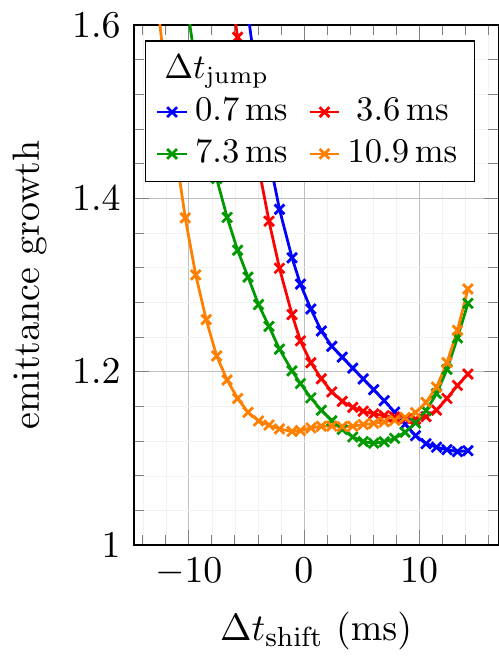}
            \caption{$\Delta\gamma_{\mathrm{t,up}}=1.5$, $\Delta\gamma_{\mathrm{t,down}}=0.5$}
        \end{subfigure}
        
        \begin{subfigure}[t]{0.32\textwidth}
        \centering
            \includegraphics{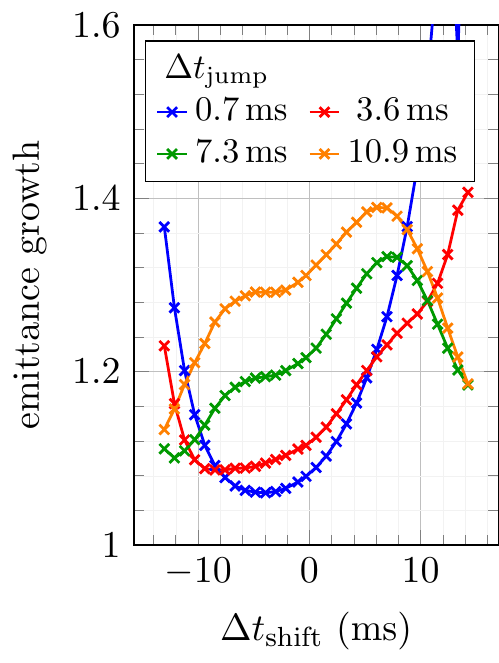}
            \caption{$\Delta\gamma_{\mathrm{t,up}}=0.5$, $\Delta\gamma_{\mathrm{t,down}}=1.0$}
        \end{subfigure}
        \hfill
        \begin{subfigure}[t]{0.32\textwidth}
            \centering
            \includegraphics{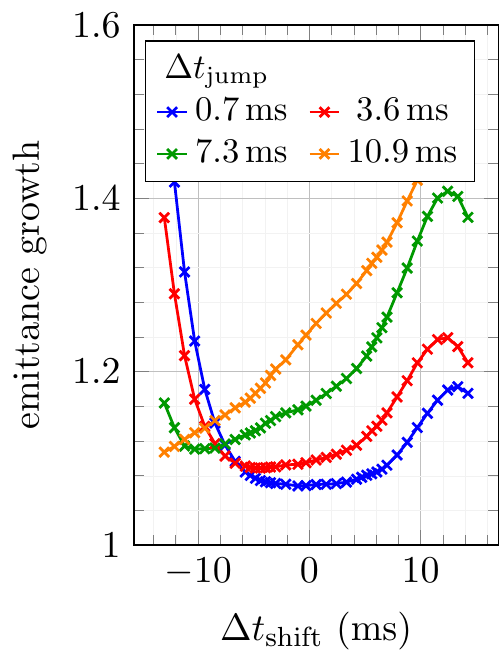}
            \caption{$\Delta\gamma_{\mathrm{t,up}}=\Delta\gamma_{\mathrm{t,down}}=1.0$}
        \end{subfigure}
        \hfill
        \begin{subfigure}[t]{0.32\textwidth}
            \centering
            \includegraphics{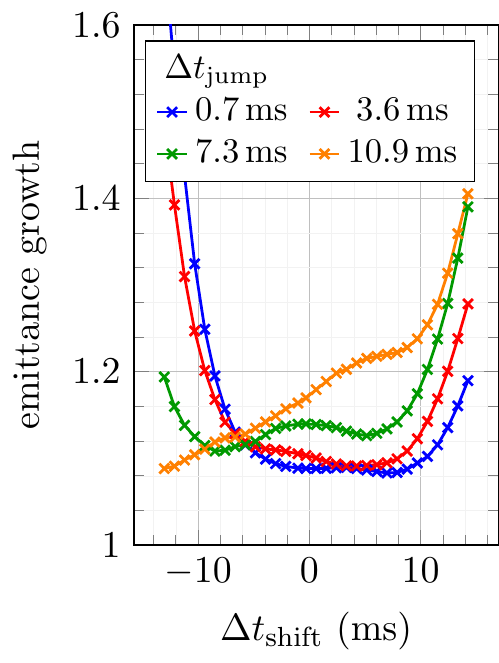}
            \caption{$\Delta\gamma_{\mathrm{t,up}}=1.5$, $\Delta\gamma_{\mathrm{t,down}}=1.0$}
        \end{subfigure}
        
        \begin{subfigure}[t]{0.32\textwidth}
        \centering
            \includegraphics{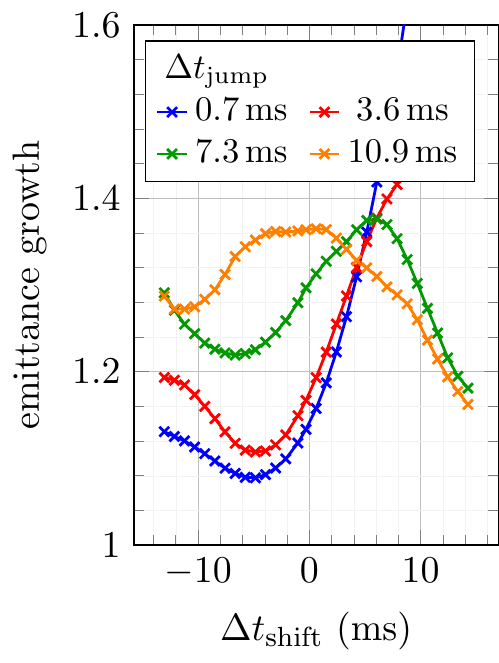}
            \caption{$\Delta\gamma_{\mathrm{t,up}}=0.5$, $\Delta\gamma_{\mathrm{t,down}}=1.5$}
        \end{subfigure}
        \hfill
        \begin{subfigure}[t]{0.32\textwidth}
            \centering
            \includegraphics{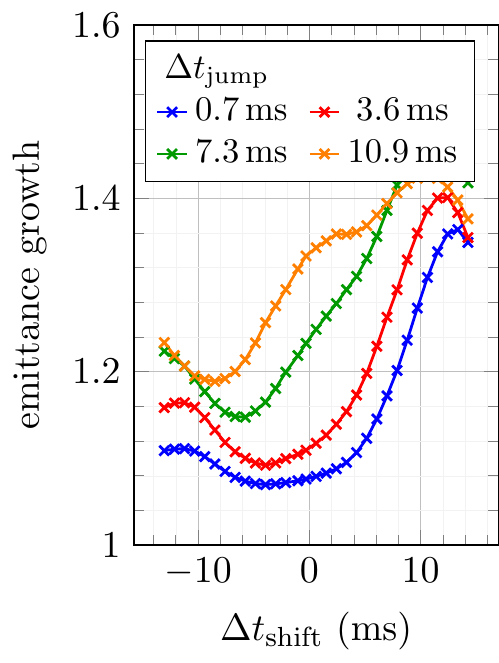}
            \caption{$\Delta\gamma_{\mathrm{t,up}}=1.0$, $\Delta\gamma_{\mathrm{t,down}}=1.5$}
        \end{subfigure}
        \hfill
        \begin{subfigure}[t]{0.32\textwidth}
            \centering
            \includegraphics{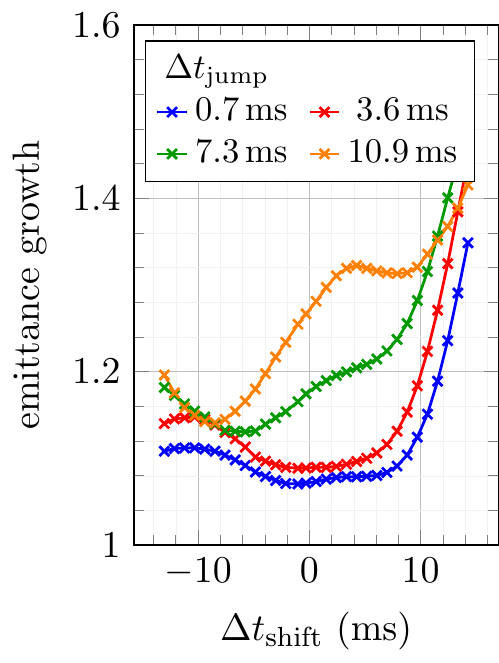}
            \caption{$\Delta\gamma_{\mathrm{t,up}}=\Delta\gamma_{\mathrm{t,down}}=1.5$}
        \end{subfigure}
        \caption{Emittance growth as function of different jump parameters}
        \label{fig:emittance_growth_jump}
    \end{figure}

\begin{figure}
        \centering
        \begin{subfigure}[t]{0.32\textwidth}
        \centering
            \includegraphics{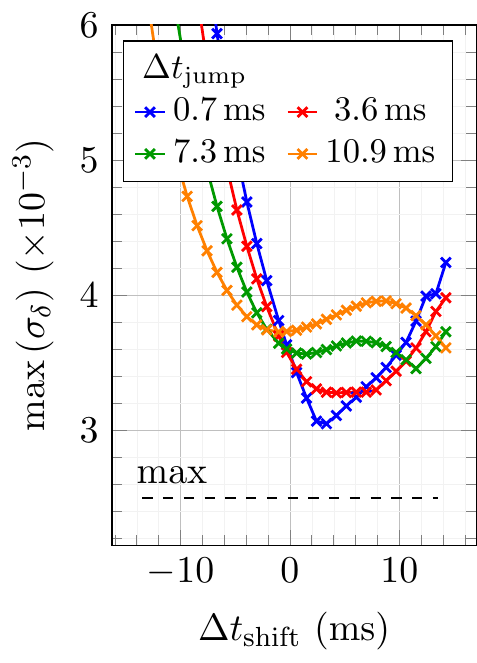}
            \caption{$\Delta\gamma_{\mathrm{t,up}}=\Delta\gamma_{\mathrm{t,down}}=0.5$}
        \end{subfigure}
        \hfill
        \begin{subfigure}[t]{0.32\textwidth}
            \centering
            \includegraphics{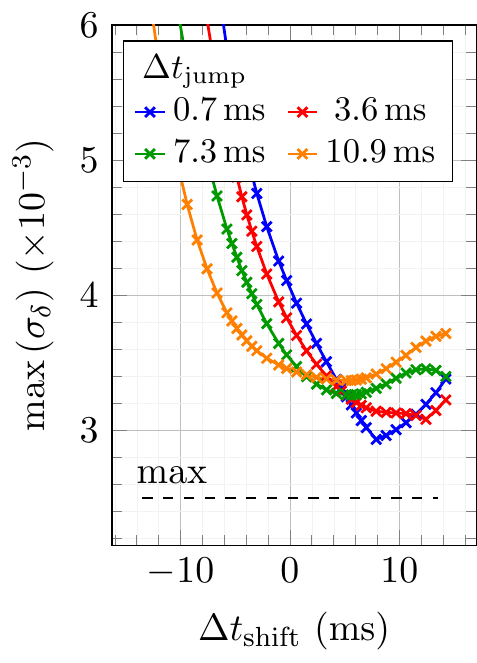}
            \caption{$\Delta\gamma_{\mathrm{t,up}}=1.0$, $\Delta\gamma_{\mathrm{t,down}}=0.5$}
        \end{subfigure}
        \hfill
        \begin{subfigure}[t]{0.32\textwidth}
            \centering
            \includegraphics{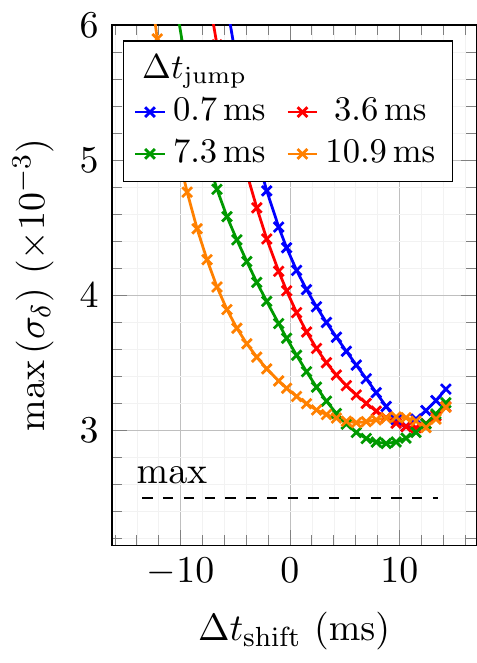}
            \caption{$\Delta\gamma_{\mathrm{t,up}}=1.5$, $\Delta\gamma_{\mathrm{t,down}}=0.5$}
        \end{subfigure}
        
        \begin{subfigure}[t]{0.32\textwidth}
        \centering
            \includegraphics{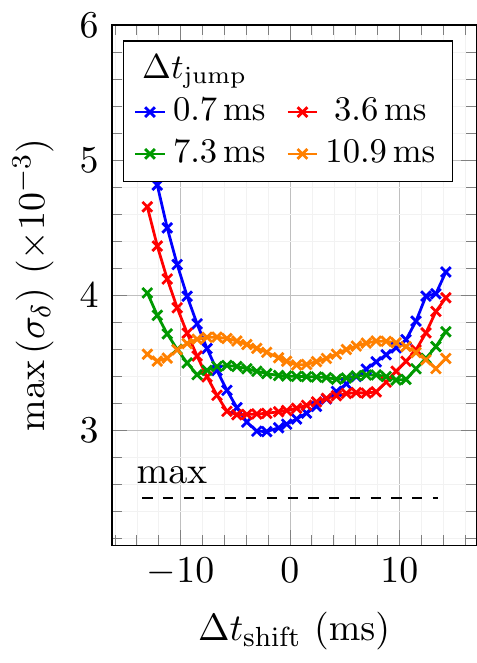}
            \caption{$\Delta\gamma_{\mathrm{t,up}}=0.5$, $\Delta\gamma_{\mathrm{t,down}}=1.0$}
        \end{subfigure}
        \hfill
        \begin{subfigure}[t]{0.32\textwidth}
            \centering
            \includegraphics{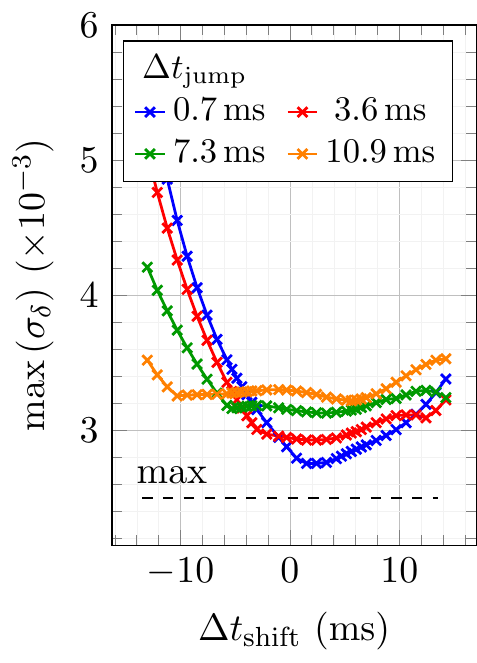}
            \caption{$\Delta\gamma_{\mathrm{t,up}}=\Delta\gamma_{\mathrm{t,down}}=1.0$}
        \end{subfigure}
        \hfill
        \begin{subfigure}[t]{0.32\textwidth}
            \centering
            \includegraphics{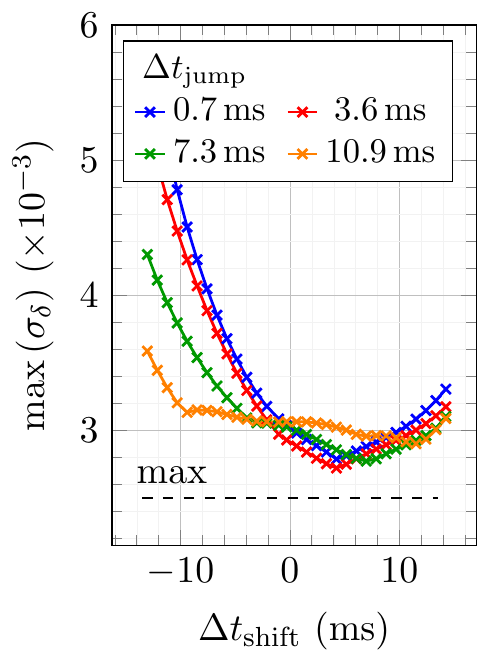}
            \caption{$\Delta\gamma_{\mathrm{t,up}}=1.5$, $\Delta\gamma_{\mathrm{t,down}}=1.0$}
        \end{subfigure}
        
        \begin{subfigure}[t]{0.32\textwidth}
        \centering
            \includegraphics{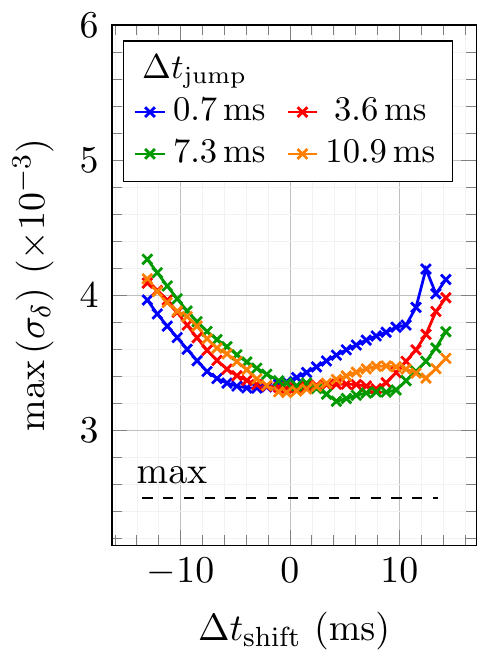}
            \caption{$\Delta\gamma_{\mathrm{t,up}}=0.5$, $\Delta\gamma_{\mathrm{t,down}}=1.5$}
        \end{subfigure}
        \hfill
        \begin{subfigure}[t]{0.32\textwidth}
            \centering
            \includegraphics{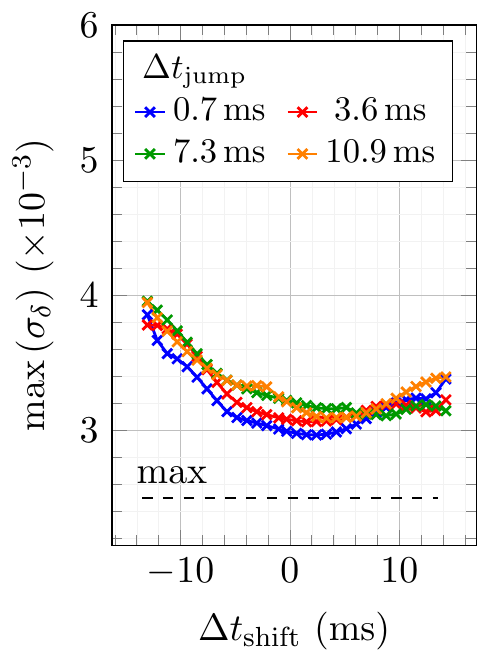}
            \caption{$\Delta\gamma_{\mathrm{t,up}}=1.0$, $\Delta\gamma_{\mathrm{t,down}}=1.5$}
        \end{subfigure}
        \hfill
        \begin{subfigure}[t]{0.32\textwidth}
            \centering
            \includegraphics{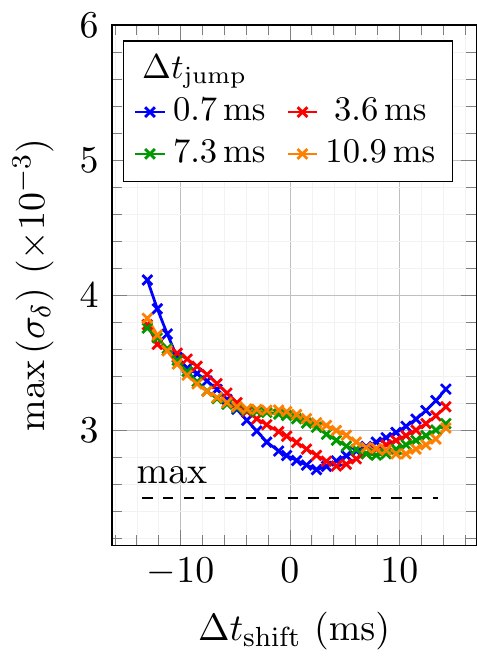}
            \caption{$\Delta\gamma_{\mathrm{t,up}}=\Delta\gamma_{\mathrm{t,down}}=1.5$}
        \end{subfigure}
        \caption{Maximum momentum spread as function of different jump parameters}
        \label{fig:maximum_momentum_spread_jump}
    \end{figure}

    \begin{figure}
        \centering
        \begin{subfigure}[t]{0.32\textwidth}
        \centering
            \includegraphics{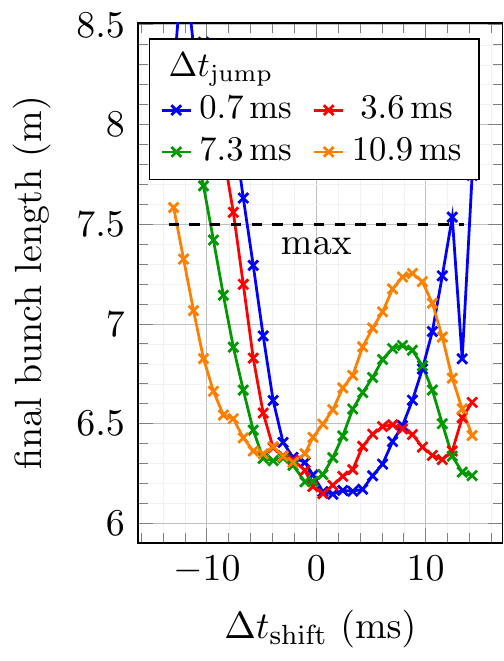}
            \caption{$\Delta\gamma_{\mathrm{t,up}}=\Delta\gamma_{\mathrm{t,down}}=0.5$}
        \end{subfigure}
        \hfill
        \begin{subfigure}[t]{0.32\textwidth}
            \centering
            \includegraphics{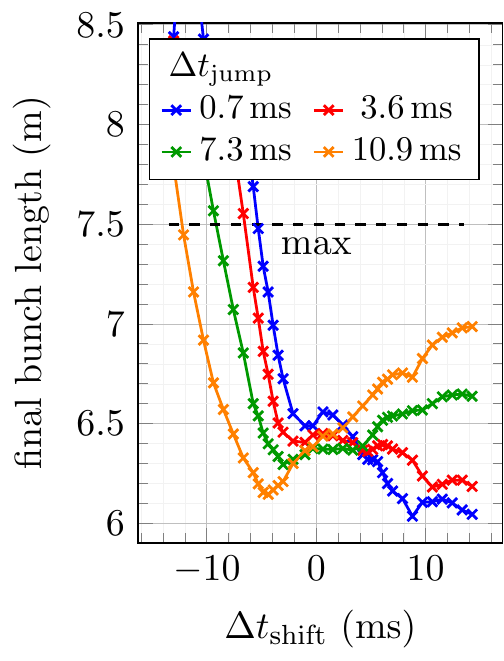}
            \caption{$\Delta\gamma_{\mathrm{t,up}}=1.0$, $\Delta\gamma_{\mathrm{t,down}}=0.5$}
        \end{subfigure}
        \hfill
        \begin{subfigure}[t]{0.32\textwidth}
            \centering
            \includegraphics{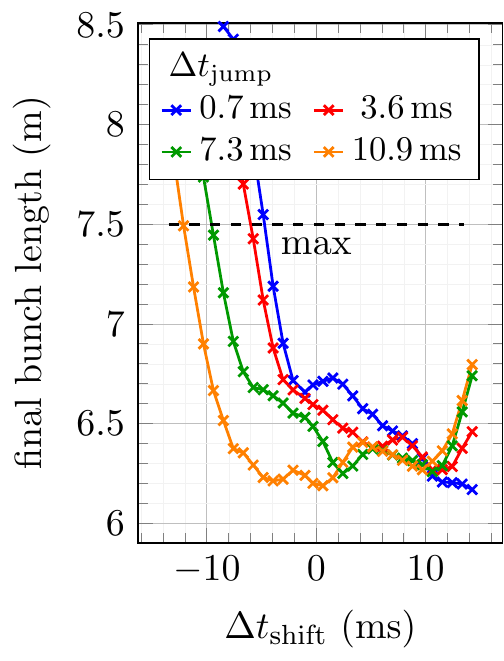}
            \caption{$\Delta\gamma_{\mathrm{t,up}}=1.5$, $\Delta\gamma_{\mathrm{t,down}}=0.5$}
        \end{subfigure}
        
        \begin{subfigure}[t]{0.32\textwidth}
        \centering
            \includegraphics{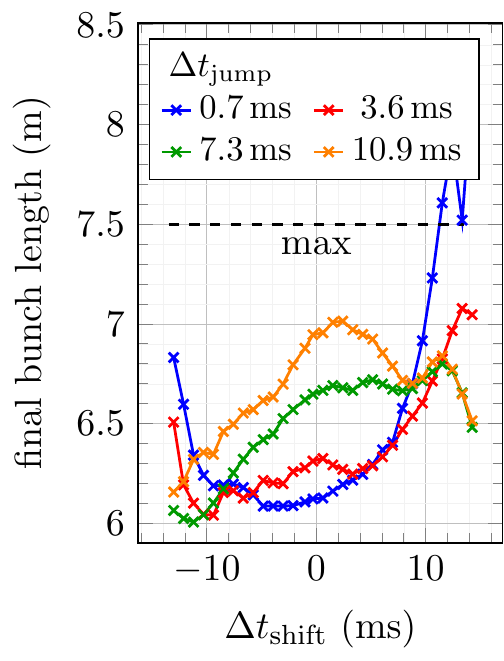}
            \caption{$\Delta\gamma_{\mathrm{t,up}}=0.5$, $\Delta\gamma_{\mathrm{t,down}}=1.0$}
        \end{subfigure}
        \hfill
        \begin{subfigure}[t]{0.32\textwidth}
            \centering
            \includegraphics{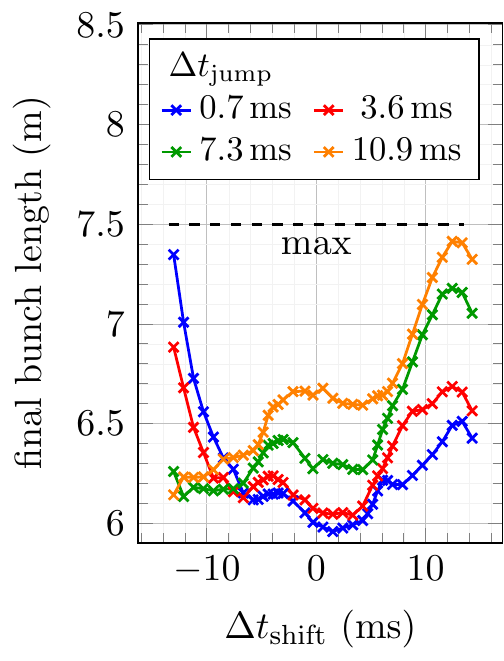}
            \caption{$\Delta\gamma_{\mathrm{t,up}}=\Delta\gamma_{\mathrm{t,down}}=1.0$}
        \end{subfigure}
        \hfill
        \begin{subfigure}[t]{0.32\textwidth}
            \centering
            \includegraphics{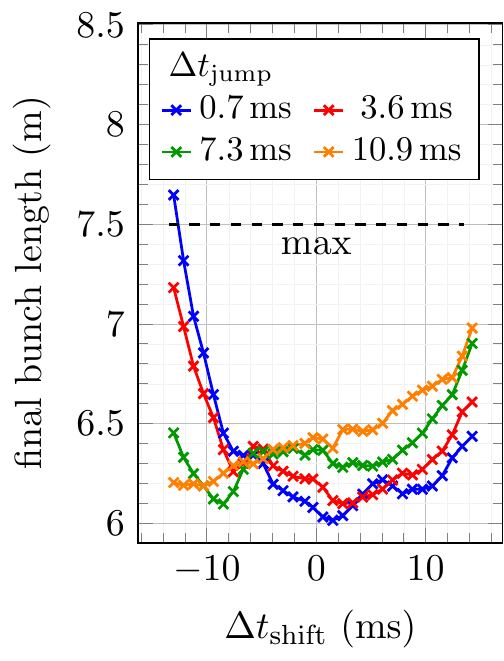}
            \caption{$\Delta\gamma_{\mathrm{t,up}}=1.5$, $\Delta\gamma_{\mathrm{t,down}}=1.0$}
        \end{subfigure}
        
        \begin{subfigure}[t]{0.32\textwidth}
        \centering
            \includegraphics{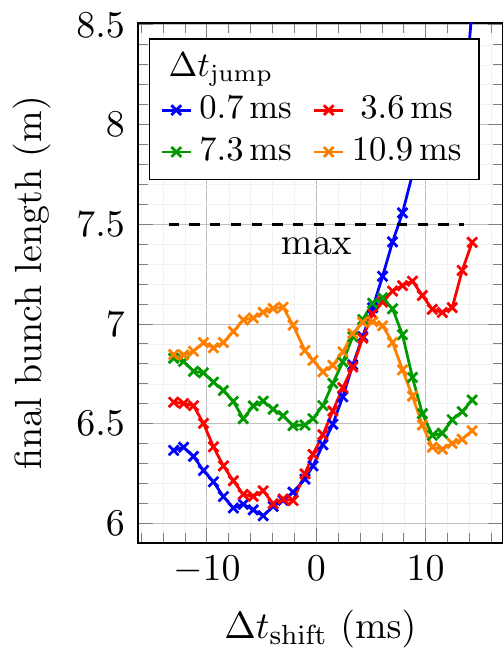}
            \caption{$\Delta\gamma_{\mathrm{t,up}}=0.5$, $\Delta\gamma_{\mathrm{t,down}}=1.5$}
        \end{subfigure}
        \hfill
        \begin{subfigure}[t]{0.32\textwidth}
            \centering
            \includegraphics{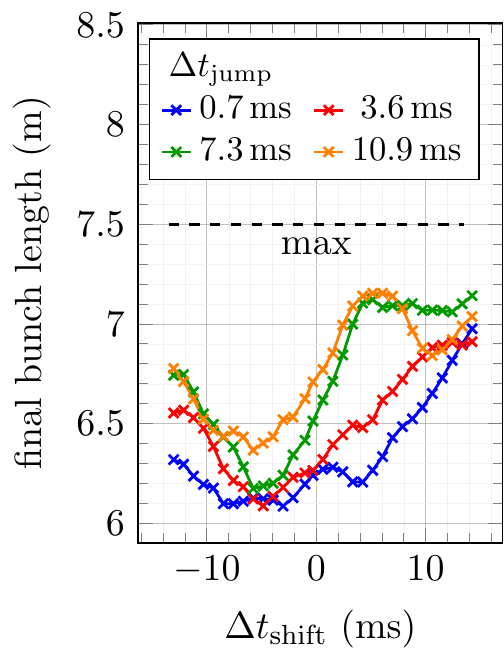}
            \caption{$\Delta\gamma_{\mathrm{t,up}}=1.0$, $\Delta\gamma_{\mathrm{t,down}}=1.5$}
        \end{subfigure}
        \hfill
        \begin{subfigure}[t]{0.32\textwidth}
            \centering
            \includegraphics{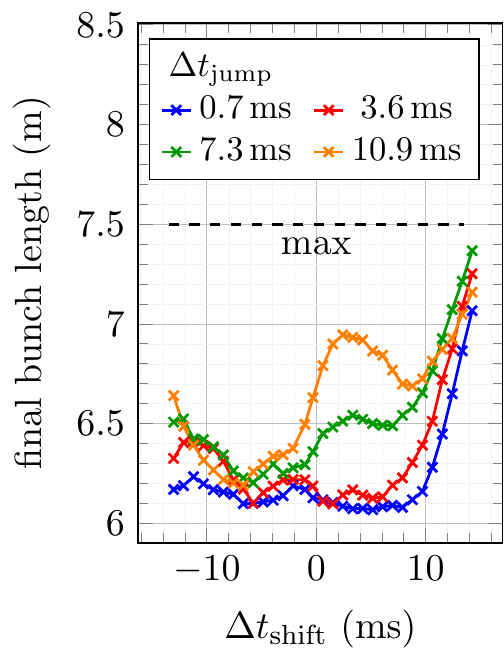}
            \caption{$\Delta\gamma_{\mathrm{t,up}}=\Delta\gamma_{\mathrm{t,down}}=1.5$}
        \end{subfigure}
        \caption{Final bunch length as function of different jump parameters}
        \label{fig:final_bunch_length_jump}
    \end{figure}
    For fixed $\Delta t_{\mathrm{slope}}=\SI{20.1}{\milli\second}$ the emittance growth $\varepsilon_{\mathrm{z}}/\varepsilon_{\mathrm{z,0}}$ variing the other parameters is plotted in Fig.~\ref{fig:emittance_growth_jump}. It can be seen that the emittance growth is always minimal for the smallest. The minimum in all simulations is reached for $\Delta\gamma_{\mathrm{t,up}}=2\cdot\Delta\gamma_{\mathrm{t,down}}=1$ and a jump shift of \SI{0.7}{\milli\second}. This asymmetric behavior avoids a missmatch in beam size due to the change of the focusing properties of space charge at transition. Below transition the equilibrium bunch length with space charge is larger than without and above transition it is the other way around. Therefore, a symmetric jump leads to quadrupolar oscillations, whereas with an asymmetric one it is possible to jump back to a matching equilibrium bunch length (cf. Figs 3 \& 4 in~\cite{MetralTRANSITIONCROSSING}). However, the simulations with these optimal jump parameters for minimal emittance growth, are not the optimum with regard to the maximum momentum spread (see Fig.~\ref{fig:maximum_momentum_spread_jump}). There, the symmetric jumps with $\Delta\gamma_{\mathrm{t,up}}=2\cdot\Delta\gamma_{\mathrm{t,down}}=1$ and $\Delta\gamma_{\mathrm{t,up}}=\cdot\Delta\gamma_{\mathrm{t,down}}=1.5$ result in a lower maximum momentum spread near to the desired maximum value (dashed black line in the plots). Last but not least the minimum of the final bunch length (shown in Fig.~\ref{fig:final_bunch_length_jump}) justifies the symmetric jump with $\Delta\gamma_{\mathrm{t,up}}=\cdot\Delta\gamma_{\mathrm{t,down}}=1$ as the optimum choice with a emittance growth around \SI{7}{\percent}, a bunch length below \SI{6}{\metre} and a maximum momentum spread of $0.027\times10^{-3}$. Note that the final bunch length of almost all simulations is below the maximum tolerable value.

\begin{figure}[htp]
        \centering
        \includegraphics{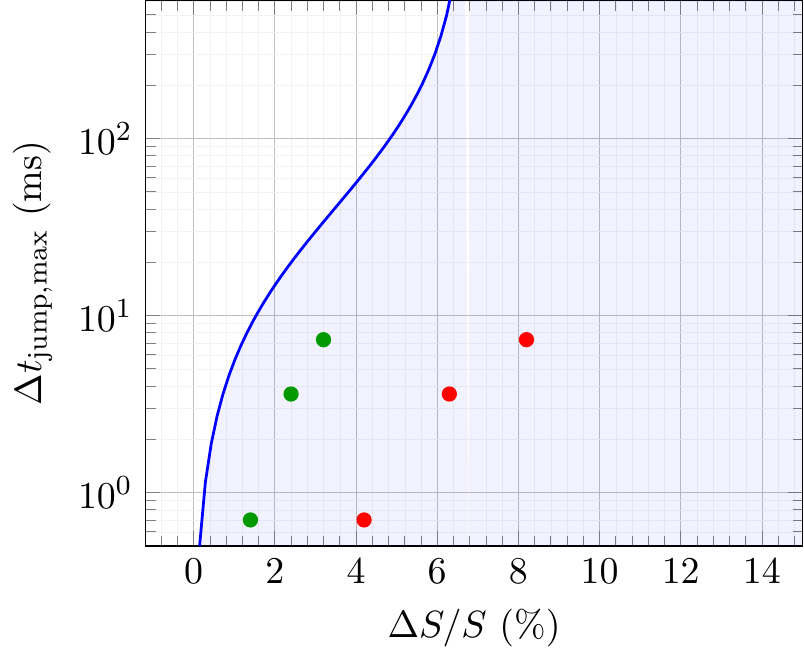}
        \caption{Maximum jump length for a maximum allowable fractional growth of bunch area from theory Eq.~\ref{eq:jump_time_constraint} (blue line) and simulated jump lengths measured directly after the jump (green) and at the end of the ramp (red)}
        \label{fig:max_jump_time}
    \end{figure}

    Similar to the discussion of transition crossing without jump (cf. Sec.~\ref{sec:transition_crossing}) theoretical requirements on the $\gamma_{\mathrm{t}}$ jump can be formulated by the emittance growth due to the nonlinear Johnsen effect~\cite{Chao1999HandbookEngineering}. To compensate this chromatic nonlinear effect the minimum jump size is given by $\Delta\gamma_{\mathrm{t}}>2\dot{\gamma}T_{\mathrm{nl}}\approx0.05$ which is fulfilled in any case. The minimum speed $\left|\dot{\gamma}_{\mathrm{t}}\right|$ of the $\gamma_{\mathrm{t}}$ jump is
    \begin{equation}
        \frac{\left|\dot{\gamma}-\dot{\gamma}_{\mathrm{t}}\right|}{\dot{\gamma}}>\left(0.76\frac{S}{\Delta S}\frac{T_{\mathrm{nl}}}{T_{\mathrm{c}}}\right)^{6/5}\mathrm{.}
    \end{equation}
    This can be reformulated as constraint on $\Delta t_{\mathrm{jump}}$ by
    \begin{equation}
        \Delta t_{\mathrm{jump}}<\frac{\Delta\gamma_{\mathrm{t}}}{{\dot{\gamma}\left\{1-\left(0.76\frac{S}{\Delta S}\frac{T_{\mathrm{nl}}}{T_{\mathrm{c}}}\right)^{6/5}\right\}}}
        \label{eq:jump_time_constraint}
    \end{equation}
    and plotted in Fig.~\ref{fig:max_jump_time} for $\Delta\gamma_{\mathrm{t}}=-(\Delta\gamma_{\mathrm{t,up}}+\Delta\gamma_{\mathrm{t,down}})=-2$.
    
    Comparing the simulation results shown as red dots in the plot (only emittance growth at and after transition) and the theoretical limit by the nonlinear effects. The emittance growth in the simulations is much larger then the theoretical limit since nonadiabatic effects on the ramp after transition -- maybe caused oscillations due to a mismatch right after transition -- lead to much larger emittance blow-up than directly at transition with a continued growth along the ramp (cf. Fig.~\ref{fig:scenario2_jump_emittance}). 
    
    \begin{figure}
        \begin{subfigure}[t]{0.32\textwidth}
            \centering
            \includegraphics{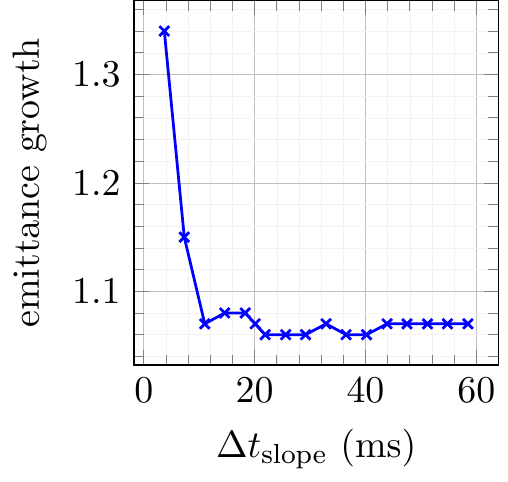}
            \caption{Emittance growth}
        \end{subfigure}
        \hfill
        \begin{subfigure}[t]{0.32\textwidth}
            \centering
            \includegraphics{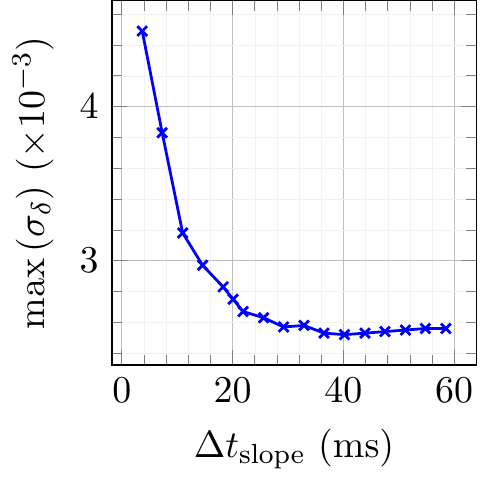}
            \caption{Maximum momentum spread}
        \end{subfigure}
        \hfill
        \begin{subfigure}[t]{0.32\textwidth}
            \centering
            \includegraphics{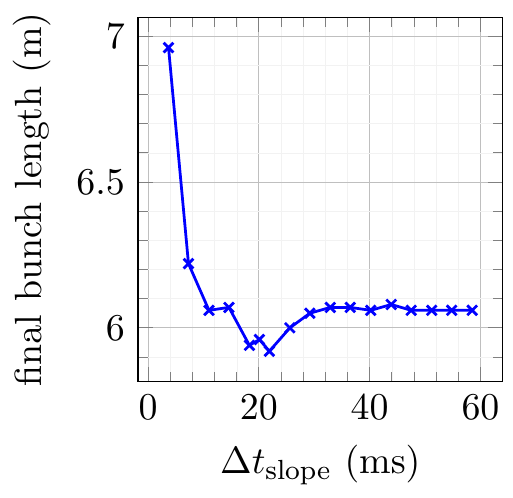}
            \caption{Final bunch length}
        \end{subfigure}
        \caption{Emittance, momentum spread and bunch length as a function of thel length of the slopes before and after the $\gamma_{\mathrm{t}}$ jump for $\Delta\gamma_{\mathrm{t,up}}=\Delta\gamma_{\mathrm{t,down}}=1.0$, $\Delta t_{\mathrm{jump}}=\SI{0.7}{\milli\second}$ and $\Delta t_{\mathrm{shift}}=\SI{0}{\second}$.}
        \label{fig:variable_slope}
    \end{figure}
    Furthermore, the already varied parameters are now chosen as fixed at the optimum and $\Delta t_{\mathrm{slope}}$ is varied. The simulations results plotted in Fig.~\ref{fig:variable_slope} show a slightly smaller maximum momentum spread at about $\Delta t_{\mathrm{slope}}=\SI{40}{\milli\second}$ then at the prior used $\Delta t_{\mathrm{slope}}=\SI{20.1}{\milli\second}$. Although this is only a small improvement, extending the length of the slope can be used to bring the maximum momentum spread near to the desired limit.
    
    Fig.~\ref{fig:scenario2_jump_statistics} summarizes the statistics for the acceleration ramp with minimum emittance growth due to transition crossing. The proton bunch has a final length of $\sigma_{\mathrm{z}}\approx\SI{6.06}{\metre}$, a momentum spread of $\sigma_\delta\approx8.1\times10^{-4}$ and a longitudinal emittance of $\varepsilon_{\mathrm{z}}\approx\SI{0.504}{\electronvolt\second}$. However, there is a emittance growth of about \SI{3}{\percent} at the beginning of the acceleration ramp which can be reduced by a more complex ramp than used for these simulations and an emittance growth of about \SI{5}{\percent} occurs during and after transition. If the number of particles in the bunch is lower, the asymmetry in the timing of the $\gamma_{\mathrm{t}}$ jump has to be adapted to the new intensity. Thus, the simulation without space charge effects but with the same jump parameters shows an larger emittance growth after transition crossing due to the missmatch at the $\gamma_{\mathrm{t}}$ jump (blue curve in Fig.~\ref{fig:scenario2_jump_emittance}).
    
     \begin{figure}
        \centering
        \begin{subfigure}[t]{0.9\textwidth}
            \centering
            \includegraphics{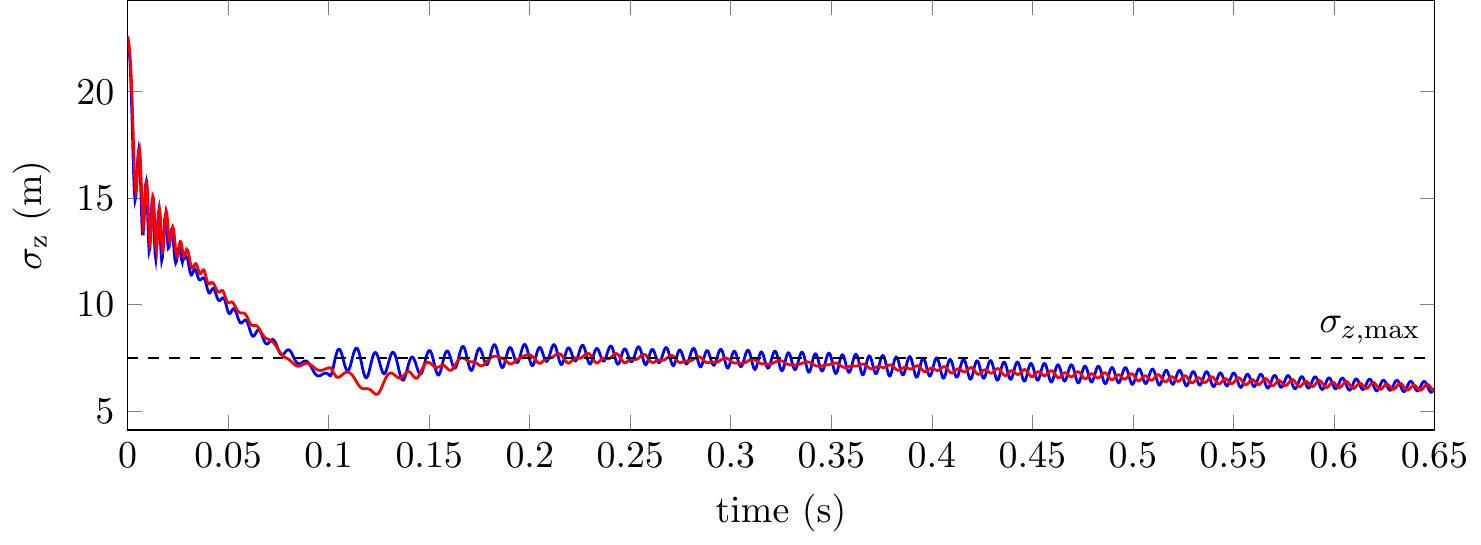}
            \caption{RMS bunch length}
           \label{fig:scenario2_jump_bunch_length}
        \end{subfigure}
        \hfill
        \begin{subfigure}[t]{0.9\textwidth}
           \centering
           \includegraphics{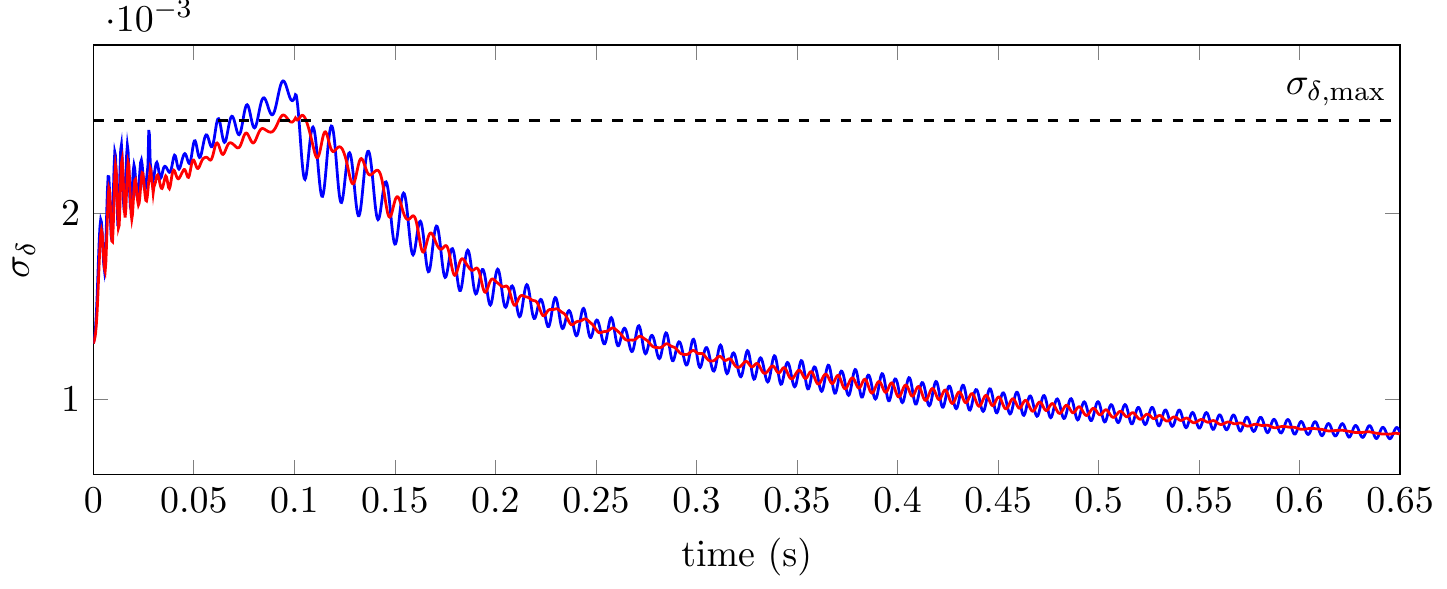}
            \caption{RMS momentum spread}
            \label{fig:scenario2_jump_momentum spread}
        \end{subfigure}
        \hfill
        \begin{subfigure}[t]{0.9\textwidth}
            \centering
            \includegraphics{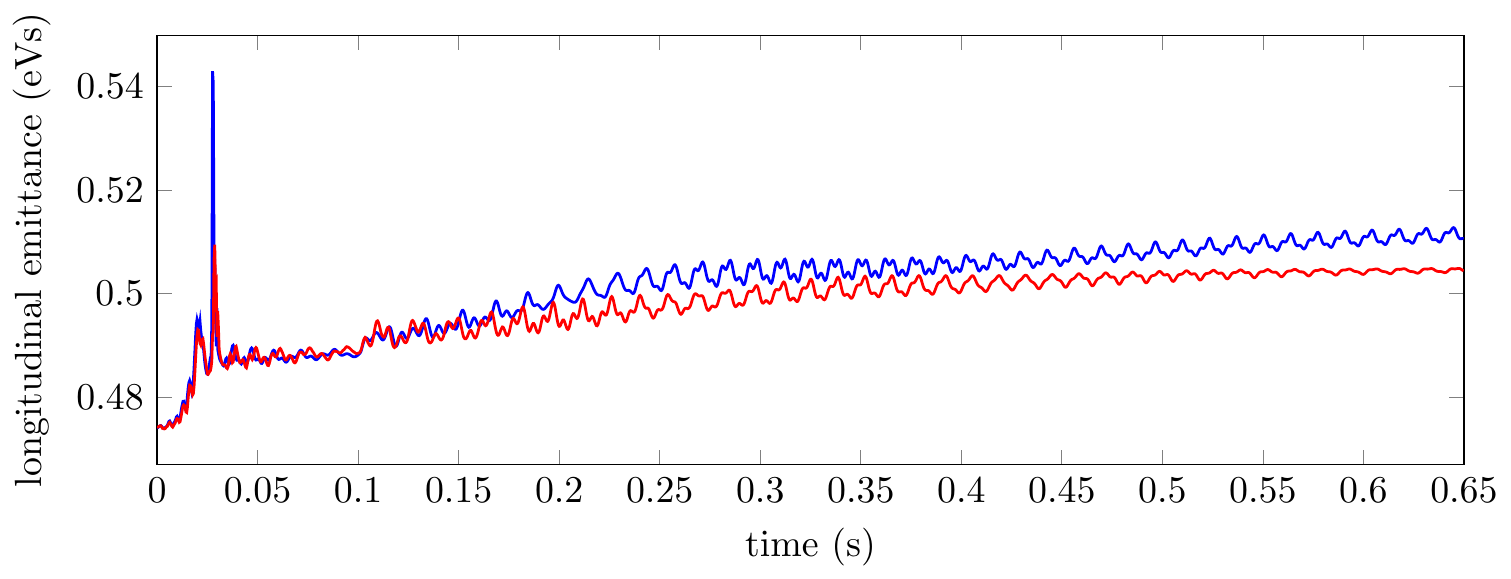}
            \caption{Longitudinal rms emittance}
            \label{fig:scenario2_jump_emittance}
        \end{subfigure}
        \caption{Statistics for scenario 2 with $\gamma_{\mathrm{t}}$ jump and modified acceleration ramp (with and without space charge in red and blue, respectively)}
        \label{fig:scenario2_jump_statistics}
    \end{figure}

\begin{figure}[htp]
        \centering
        \includegraphics{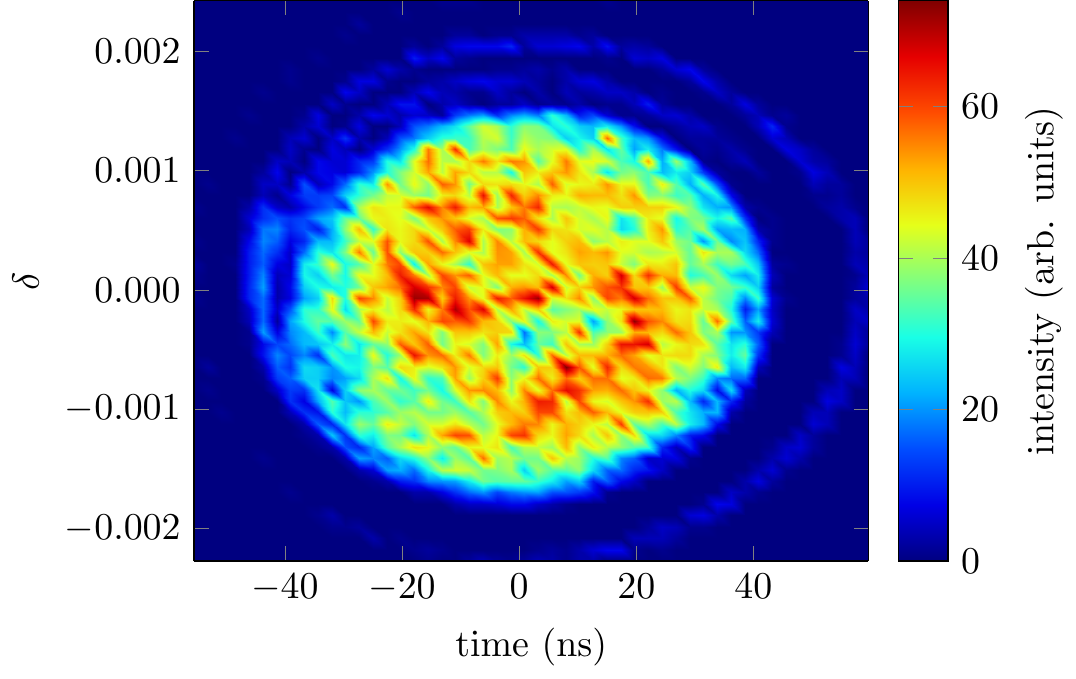}
        \caption{Phase space at the end of the acceleration ramp for scenario 2 with $\gamma_{\mathrm{t}}$ jump and with space charge for the parameters shown in Tab.~\ref{tab:best_parameters}}
        \label{fig:scenario2_phasespace}
    \end{figure}

    To conclude, the best jump parameters are summarized in Tab~\ref{tab:best_parameters} and the phase space after acceleration is plotted in Fig.~\ref{fig:scenario2_phasespace}.
    \begin{table}[thp]
        \begin{center}
        \caption{\label{tab:best_parameters} $\gamma_{\mathrm{t}}$ jump parameters with minimal emittance growth at nominal intensity of $2\times10^{13}$ particles}
        \begin{tabular}{l c c}
        \hline
        Parameter & Value (in turns) & Value in ms\\
        \hline
        $\Delta\gamma_{\mathrm{t,up}}$&1.0&-- \\
        $\Delta\gamma_{\mathrm{t,down}}$&1.0&--\\
        $\Delta t_{\mathrm{slope}}$ &11000&40.2 \\
        $\Delta t_{\mathrm{jump}}$ &200&0.7 \\
        $\Delta t_{\mathrm{shift}}$ &400&1.4 \\\hline
        \end{tabular}
        \end{center}
    \end{table}

\section{Conclusion and Outlook\label{sec:conclusion}}
The SIS100 proton cycle accelerates protons from \SI{4}{\giga\electronvolt} to \SI{29}{\giga\electronvolt}. The proposed lattice design requires taking care of the beam dynamics around transition energy. For this, two different scenarios are foreseen: Shifting transition energy during the acceleration ramp to avoid transition crossing and crossing transition with a $\gamma_{\mathrm{t}}$ jump to control the possible blow-up in phase space.

The longitudinal simulations presented in this report have shown that shifting the transition energy during the ramp -- especially if this is done smoothly -- would lead to a bunch distribution fulfilling all requirements (cf. Tab.~\ref{tab:final_beam_parameters}). For the implementation, however, the transition energy has to be increased above $\gamma_{\mathrm{t}}=40$.

For the scenario with constant $\gamma_{\mathrm{t}}=8.9$, the maximum allowed momentum spread of $\delta_{\mathrm{max}}=2\sigma_{\delta,\mathrm{max}}=5\times10^{-3}$ would be exceeded near to transition energy and a not tolerable emittance blow-up caused by space charge effects would occur. To reduce this emittance growth significantly a $\gamma_{\mathrm{t}}$ jump is introduced and the parameters of this jump were optimized at SIS100 nominal intensity of $2\times 10^{13}$ particles per bunch (cf. Tab.~\ref{tab:best_parameters}). This optimization results in a $\gamma_{\mathrm{t}}$ jump with below \SI{7}{\percent} emittance growth due to transition crossing fulfilling the requirements on bunch length and momentum spread (cf. Tab.~\ref{tab:final_beam_parameters}). 
Furthermore, the simulations indicated that the most critical requirement is the maximum momentum spread along the ramp. If this could be relaxed, the $\gamma_{\mathrm{t}}$ jump does not have to be as fast as suggested for example in~\cite{Aumon2014TransitionOperation}. However, only longitudinal space charge effects are included in the simulations so far. Transverse effects could also require a faster jump and have to be studied in the future.

\bibliography{references}

\begin{thebibliography}{10}

\bibitem{Kornilov2013OverviewCycles}
V.~Kornilov, O.~Boine-Frankenheim, and D.~Ondreka, ``{Overview of the
  Longitudinal Beam Dynamics for the SIS100 Proton Cycles},'' {\em SIS Project
  Note, WP 2.8.1 SIS100 Beam Dynamics}, 2013.

\bibitem{Bar2008FAIRSIS100}
R.~B{\"{a}}r, U.~Blell, O.~Boine-Frankenheim, L.~Bozyk, K.~Blasche,
  J.~Falenski, E.~Fischer, E.~Floch, G.~Franchetti, B.~Franczak, P.~Forck,
  O.~Gumenyuk, I.~Hofmann, P.~H{\"{u}}lsmann, M.~Kauschke, M.~Kirk,
  H.~Klingbeil, H.~G. K{\"{o}}nig, A.~D. Kovalenko, P.~Kowina, A.~Kr{\"{a}}mer,
  D.~Kr{\"{a}}mer, M.~Kumm, U.~Laier, M.~Mehler, J.~P. Meier, G.~Moritz,
  P.~Moritz, C.~M{\"{u}}hle, K.~P. Ningel, C.~Omet, I.~Pschorn, N.~Pyka,
  H.~Ramakers, P.~Schnizer, G.~Schreiber, C.~Schroeder, M.~Schwickert, Y.~Shim,
  M.~Sitko, B.~Skoczen, S.~Sorge, P.~Spiller, J.~Stadlmann, A.~Stafiniak,
  K.~Sugita, B.~Weckenmann, and H.~Welker, ``{FAIR Technical Design Report
  SIS100},'' tech. rep., GSI Helmholtzzentrum f{\"{u}}r Schwerionenforschung
  GmbH, Darmstadt, Germany, 2008.

\bibitem{Sorge2012BeamEffects}
S.~Sorge, ``{Beam Dynamics Study Concerning SIS-100 Proton Operation Including
  Space Charge Effects},'' in {\em ICAP2012}, 2012.

\bibitem{Yuan2021RFSIS-100}
Y.-S. Yuan, O.~Boine-Frankenheim, T.~Egenolf, and V.~Kornilov, ``{RF
  manipulations of high-intensity hadron beams in SIS-100},'' tech. rep., GSI
  Helmholtzzentrum f{\"{u}}r Schwerionenforschung GmbH, Darmstadt, Germany,
  2021.

\bibitem{Ng2006PhysicsInstabilities}
K.~Y. Ng, {\em {Physics of Intensity Dependent Beam Instabilities}}.
\newblock Singapore, New Jersey, London, Hong Kong: World Scientific, 2006.

\bibitem{Lee1999AcceleratorPhysics}
S.~Y. Lee, {\em {Accelerator Physics}}.
\newblock Singapore, New Jersey, London, Hong Kong: World Scientific, 1999.

\bibitem{Gilardoni2010Fifty1}
S.~Gilardoni, D.~Manglunki, and {European Organization for Nuclear Research.},
  {\em {Fifty years of the CERN proton synchrotron. Vol. 1}}.
\newblock CERN, 2010.

\bibitem{Johnsen1956EffectsTransition}
K.~Johnsen, ``{Effects of Non-linearities on the Phase Transition},'' in {\em
  HEAAC 1956}, 1956.

\bibitem{Chao1999HandbookEngineering}
A.~W. Chao and M.~Tigner, {\em {Handbook of Accelerator Physics and
  Engineering}}.
\newblock Singapore: World Scientific Publishing Co. Pte. Ltd., 1999.

\bibitem{Lens2013StabilitySynchrotrons}
D.~Lens and H.~Klingbeil, ``{Stability of longitudinal bunch length feedback
  for heavy-ion synchrotrons},'' {\em Physical Review Special Topics -
  Accelerators and Beams}, vol.~16, 3 2013.

\bibitem{Sorge2018SimulationOperation}
S.~Sorge, ``{Simulation study on beam loss in the alpha bucket regime during
  SIS-100 proton operation},'' {\em Nuclear Instruments and Methods in Physics
  Research, Section A: Accelerators, Spectrometers, Detectors and Associated
  Equipment}, vol.~882, pp.~129--137, 2 2018.

\bibitem{MetralTRANSITIONCROSSING}
E.~M{\'{e}}tral and D.~M{\"{o}}hl, ``{Transition Crossing},'' tech. rep.

\bibitem{Aumon2014TransitionOperation}
S.~Aumon, D.~Ondreka, S.~Sorge, and K.~Gross, ``{Transition energy crossing in
  the future FAIR SIS-100 for proton operation},'' in {\em 5th International
  Particle Accelerator Conference IPAC2014}, JACoW, 2014.

\end{thebibliography}
\end{document}